\documentclass[trackchanges]{aastex701}

\usepackage{amsmath,amssymb}

\usepackage{subfig}
\newcommand{\pd}[2]{\frac{\partial #1}{\partial #2}}
\usepackage{enumitem}

\usepackage{etoolbox}
\AfterEndEnvironment{subequations}{\par\noindent\ignorespaces}
\usepackage{changepage}
\usepackage{multirow}
\usepackage{comment}

\begin{document}

\title{Chemical transport by weakly nonlinear internal gravity waves in stars}

\author[orcid=0000-0003-1863-8945]{Yifeng Mao}
\affiliation{Scripps Institution of Oceanography, University of California, San Diego, La Jolla, CA, USA}
\email[show]{yifeng\_mao@brown.edu}

\author[orcid=0000-0002-7635-9728]{Daniel Lecoanet}
\affiliation{Center for Computational Astrophysics, Flatiron Institute, New York, NY, USA}
\affiliation{Department of Engineering Sciences and Applied Mathematics Northwestern University, EvanstonIL, USA}
\affiliation{Center for Interdisciplinary Exploration and Research in Astrophysics (CIERA), Northwestern University, Evanston, IL, USA}
\email[]{}

\begin{abstract}

While internal gravity waves (IGWs) are known to transport chemicals in the radiative zones of stars, substantial uncertainty remains regarding the magnitude and physical mechanisms of this transport. Most previous studies have relied on heuristic theories or numerical simulations that are difficult to extrapolate to stellar parameters. 
In this work, we present the first rigorous asymptotic calculation of (passive) chemical transport by IGWs, in the limit of small wave amplitude.
The net vertical transport induced by a coherent packet of waves is found to be proportional to the horizontal mean of $(w^S+w^E)^2$, where $w^S$ and $w^E$ are the vertical Stokes and Eulerian mean velocities. The transport scales as the fourth power of wave amplitude and quadratically with time, and we verify these scalings with numerical simulations. In the weakly damped regime, the dominant transport coefficients scale as $\omega^{-2}\kappa^2$, where $\omega$ is the wave frequency and $\kappa$ is the radiative diffusivity. Because the transport can be positive or negative depending on the wave phases, the effective mixing induced by the superposition of coherent wave packets is expected to scale with the eighth power of wave amplitude. These results provide a theoretical foundation for making accurate predictions of IGW--induced chemical transport in stellar evolution models.


\end{abstract}

\keywords{}

\section{Introduction} \label{sec:intro}

Because many of the most common elements are primarily synthesized in stellar cores \citep{Nomoto2013}, stellar evolution and chemical transport in stars play important roles in determining the chemical evolution of the universe \citep[e.g.,][]{Kaiser2020}.
Different regions of stars have different dominant chemical transport mechanisms \citep{Salaris2017}.
Bulk fluid mixing is known to homogenize chemical species in convection zones over stellar evolutionary timescales, except in the latest and shortest stages of a star's life \citep{Andrassy2020}.
At the boundaries of convective and radiative zones, different mechanisms including convective overshoot and convective penetration have been invoked to explain observations and interpret simulations \citep[see][for a review]{Anders2023}.
However, chemical transport in radiative zones remain a major uncertainty in stellar evolution theory.
Several different proposed mechanisms include, Eddington-Sweet circulation \citep[e.g.,][]{Maeder1998}, horizontal and vertical shear mixing \citep{Garaud2024}, MHD turbulence driven by radial shear \citep{Spruit2002}, and wave mixing \citep{Rogers2017}.
It is still an open question how to best parameterize these mechanisms in stellar evolution models, and which may be most important for different types of stars.

Asteroseismic observations from space-based photometry missions, including CoRoT \citep{Baglin2006, Auvergne2009}, Kepler \citep{Borucki2010}, and TESS \citep{Ricker2015}, have revolutionized our understanding of stellar structure and evolution \citep{Aerts2010}.
\citet{Pedersen2021} analyzed internal gravity wave period-spacing data for 26 slowly-pulsating B stars.
They compared these data to stellar evolution models with different chemical mixing properties.
The best-fit models require both convective boundary mixing, as well as mixing in the radiative zones of the stars.
The $\beta$ Cep star HD 192575 was also inferred to have both convective boundary and radiative zone mixing in \citet{Burssens2023}.
In addition, chemical transport can be constrained by spectroscopic surveys which measure the surface abundances of different chemical species in massive stars \citep[e.g.,][]{Hunter2009,Gebruers2021}.
Often, the surface abundances only match stellar models which include some envelope mixing \citep[e.g.][]{Maeder2014}.
Recent works have specifically tested wave mixing as a mechanism to explain surface abundance patterns \citep{Mombarg2025,Brinkman2025}.

These works highlight the need for accurate a priori models of radiative zone mixing.
In this work we will focus on the wave mixing mechanism.
At high enough amplitude, the shear generated by internal gravity waves (IGWs) can overcome the stabilizing stratification, driving shear instabilities \citep{GarciaLopez1991,Zahn1992}.
IGWs could be particularly susceptible to shear instabilities in stellar interiors, where the thermal diffusivity $\kappa$ is much larger than the viscosity $\nu$ \citep[i.e., Prandtl number $Pr=\nu/\kappa\ll1$, see][]{Cope2020}.
These strongly nonlinear waves are theorized to produce diffusive mixing with diffusivity $D_{w}\sim u^2 \kappa$, where $u$ is the IGW amplitude.
In contrast, \citet{Jermyn2022} calculates a wave mixing diffusivity for weakly nonlinear waves.
He assumes the Stokes drift from different waves incoherently transport chemical species, leading to a mixing diffusivity $D_{w}\sim u^4 D^2$, where $D$ is the microphysical chemical diffusivity.
More recently, \citet{mathis2026understanding} applied the Osborn–Cox closure model to low-frequency IGWs and obtained an effective diffusivity $D_w\sim \kappa u^2$, consistent with the \citet{Zahn1992} shear-mixing prescription.
While there is substantial uncertainty in the amplitude of stochastically-excited IGWs in stars \citep[e.g.,][]{Horst2020,LeSaux2023,Ratnasingam2023,Anders2023b,Thompson2024}, current simulations do not appear to show IGW-driven shear instabilities (especially since $Pr\sim 1$ in most simulations).

These theoretical uncertainties have led to numerical investigations of wave mixing.
The first numerical study was \citet{Rogers2017}, which advected passive tracer particles by waves generated by convection in a 2D anelastic simulation of a $3M_\odot$ zero-age main sequence (ZAMS) star.
They find the particles radially diffuse in time, and measure a wave mixing diffusion coefficient $D\sim u^2$, in agreement with \citet{GarciaLopez1991}, despite a lack of obvious shear instabilities.
A similar approach has subsequently been applied to other stellar types and rotation rates \citep{Varghese2023, Varghese2024,Varghese2025}, with qualitatively similar results.
\citet{Higl2021} also evolved passive tracer particles in a 2D pseudo-incompressible simulation of a $3M_\odot$ ZAMS star.
They found the wave diffusivity decreased as they increased the simulation resolution.
\citet{Herwig2023} solve an advection equation for a passive tracer field, and determine how it is mixed by waves right outside of the convection zone of a $25M_\odot$ early-main sequence star.
Their results also appear to be in line with \citet{GarciaLopez1991}, although again there are not obvious shear instabilities in their radiative zone.
Finally, \citet{Morton2025} pointed out that particle tracking algorithms are susceptible to numerical errors related to clumping which can artificially enhance measured diffusion coefficients.
However, the recent work of \citet{Varghese2025} shows that this clumping mechanism process does not play a role in their particle tracking calculation.
In summary, there is no firm consensus on the magnitude, functional dependence, or physical mechanism behind IGW-driven wave mixing in stars.

In this paper, we address these uncertainties by studying an idealized wave mixing problem.
We consider the superposition of low-amplitude waves propagating in a cartesian domain in the Boussinesq approximation (Section~\ref{sec:setup}).
While previous numerical studies calculated mixing by IGWs self-consistently generated by convection, we exploit the simplicity of our setup to solve the problem both analytically and numerically.
Using the wave amplitude as an asymptotic expansion parameter, we develop a multiscale asymptotic expansion and derive the evolution of a passive tracer field due to advection by these (weakly) nonlinear waves (Section~\ref{sec:result}).
We then verify these calculations by comparison to numerical simulations in which a small number of upward-propagating waves are forced from the bottom boundary for several wave periods (Section~\ref{sec:sims}).
We find that the horizontal-mean passive tracer field evolves non-diffusively, with an amplitude that scales like $\kappa^2 u^4$. 
In Section~\ref{sec:coeffient} we determine which waves produce the most mixing, and in Section~\ref{sec:discussion} we discuss the implications of this non-diffusive wave mixing for stellar evolution.

\section{Problem setup} \label{sec:setup}


\subsection{The governing equations}

For our idealized setup, we consider a 2D Cartesian model with a passive tracer field in a non-rotating, stratified medium within a uniform gravitational field $-g\mathbf{e_z}$. In the linear Boussinesq approximation when the fluid buoyancy is subject to a small variation $b=b_0+b'$ and pressure is also perturbed as $p=p_0+p'$, we obtain the following system
\begin{subequations} \label{eq:Boussq}
\begin{align}
\frac{\partial \mathbf{u}}{\partial t} + \nabla p - b \mathbf{e}_z &= -\mathbf{u} \cdot \nabla \mathbf{u}, \\
\frac{\partial b}{\partial t} - \kappa \nabla^2 b + N^2 w &= - \mathbf{u} \cdot \nabla b, \\
\nabla \cdot \mathbf{u} &= 0,
\end{align}
\end{subequations}
where the prime notation is dropped for simplicity, $\mathbf{u}=(u,v)$ is the velocity field, $p$ is the pressure perturbation, $b$ is the buoyancy perturbation, $\kappa$ is the radiative diffusivity, and $N^2=db_0/dz$ is the buoyancy frequency. Because the radiative diffusivity $\kappa$ is much higher than the viscosity and chemical diffusivities in stars, we only include radiative diffusivity in our equations. The transport equation for a passive chemical species $c$ is given by
\begin{equation}
\frac{\partial c}{\partial t} + \mathbf{u} \cdot \nabla c = 0, \quad c(\mathbf{x},t=0)=c_0(\mathbf{x}), \label{eq:transport}
\end{equation}
which describes the evolution of the passive tracer over time due to advection by the fluid flow.

\subsection{Simulation setup} \label{sec:sim_setup}

In addition to the analytic calculations described below, we also solve equations \eqref{eq:Boussq} \& \eqref{eq:transport} numerically with the Dedalus pseudo-spectral code \citep{BurnsDedalus2020}. 
The simulations use 256 Fourier modes in the horizontal $x$ direction and 512 Chebyshev modes in the vertical $z$ direction. 
The wavenumbers are selected so that the wave field is periodic in the horizontal direction.
Under our nondimensionalization (described below), the simulation domain is $(x,z)\in[0,2\pi]\times[0,4\pi]$ such that the largest horizontal wave mode has $|k_h|=1$. 
Time integration is carried out using a fourth-order semi-implicit backward differentiation formula (BDF) scheme \citep{wang2008variable}, with a time step $\Delta t =0.004\pi$.

Wave forcing is imposed at the lower boundary $z=0$ through the vertical velocity condition
\begin{align}
    w(x,z=0,t) = \sum_j a_j \cos(k_{h,j} x-\omega_j t) F(t), \quad 
    F(t) &= \frac{1}{2} \left(\tanh\frac{t-t_0}{T_0} - \tanh\frac{t-t_1}{T_0} \right), \label{eq:BC} 
\end{align}
where $a_j$, $k_{h,j}$ and $\omega_j$ denote the amplitude, wavenumber, and frequency of the imposed waves. The envelope function smoothly turns the forcing on and off, where $T_0=100$ is chosen to control the timescale of the transition, and $t_0=150\pi $ and $t_1=1200\pi$ denote the onset and termination times of the forcing. Gradually turning on and off the wave forcing keeps the wave signal close to the sum of monochromatic waves, and enables us to compare the passive tracer deviation before and after the forcing period to quantify the net transport induced by the waves.
The buoyancy field $b(x,z=0,t)$ is specified to be the exact solution from the linear system 
\begin{equation*}
    -\pd{}{t} \left( \nabla^2 w \right) + \nabla_h^2 b =0, \quad \text{or equivalently, } \quad \pd{b}{t} - \kappa \nabla^2 b + N^2 w=0,
\end{equation*}
where $\nabla_h^2$ is the horizontal Laplacian. These two forms are equivalent under the linear dispersion relation, so either can be used to prescribe the boundary condition for $b(x,z=0,t)$.

At the upper boundary $z=4\pi$, we impose $w(x,z=4\pi,t)=0$ and $b(x,z=4\pi,t)=0$.
To further absorb outgoing waves and inhibit wave reflections, we include a damping layer $G(z)$ in the upper region of the domain. While waves reflection does occur in stars, the goal of our numerical calculations is to compare with analytic calculations in which, for simplicity, we assume the waves propagate upward.
The buoyancy equation is modified as
\begin{equation}
\pd{b}{t} - \kappa \nabla^2 b + N^2 w = -b G(z), \quad G(z) = \frac{1}{2} [1 + \text{erf}((z - z_1) / \sigma_1)], \label{eq:BC_b}
\end{equation}
where $z_1=2.8\pi$ marks the onset of the damping region and $\sigma_1=1$ controls its vertical extent. This damping layer progressively reduces the wave amplitude before the waves reaches the top boundary at $z=4\pi$.

The boundary conditions are derived from the linear equations for harmonic waves, since the nonlinear solution is unknown, so we cannot use it to set boundary conditions. To incorporate nonlinear effects, we instead introduce nonlinearity only above $z=0$. It also ensures numerical stability by avoiding spurious nonlinear effects near the boundary. The governing equations are then modified to
\begin{align*}
    \frac{\partial \mathbf{u}}{\partial t} + \nabla p - b \mathbf{e}_z &= - H(z) \left( \mathbf{u} \cdot \nabla \mathbf{u} \right) , \\
    \frac{\partial b}{\partial t} - \kappa \nabla^2 b + N^2 w &= - H(z) \left( \mathbf{u} \cdot \nabla b \right),
\end{align*}
where $H(z) = \frac{1}{2} \left(\text{erf}((z - z_0) / \sigma_2) - \text{erf}((z - z_1) / \sigma_2) \right)$ is a smooth windowing function that localizes the nonlinear terms to the region $z_0<z<z_1$, where $z_0=0.4\pi$ and $z_1=2.8\pi$.

To analyze chemical transport, we solve the modified advection equation
\begin{align}
   \frac{\partial c}{\partial t} = -  H(z)\left ( \mathbf{u} \cdot \nabla c \right) , \label{eq:BC_c}
\end{align}
where $H(z)$ is the same windowing function defined earlier, which localizes transport to the interior. Since the advection term vanishes near the boundaries, no explicit boundary conditions for $c$ at $z=0$ and $z=4\pi$ are required. 
The passive tracer field is initialized as $c_0 \equiv c(z,t=0)=z^2$, and we track its evolution over time. 
In the following analysis, we focus on the central interval $z\in[0.8\pi,2\pi]$, where $c$ satisfies the advection equation.

The presented simulation results are nondimensionalized using the characteristic horizontal length scale $|k_h|^{-1}$ and time scale $N^{-1}$, leading to the transformations
\begin{equation}
    \hat{\mathbf{x}}= |k_h| \mathbf{x}, \quad \hat{t}=N t, \quad \hat{\mathbf{u}}=\frac{|k_h|}{N}\mathbf{u}, \quad \hat{p} = \frac{k_h^2}{N^2} p, \quad \hat{b}= \frac{|k_h|}{N^2} b, \quad \hat{\kappa}=\frac{k_h^2}{N}\kappa,
\end{equation}
Hats denote nondimensional quantities and are omitted hereafter for simplicity.
With these scalings, the wave system \eqref{eq:Boussq} takes a nondimensional form in which $N^2$ is eliminated. The transport equation \eqref{eq:transport}, being linear in $c$, remains scaling-invariant and therefore unchanged by the nondimensionalization.

\subsection{Net vertical transport}

In stellar structure models, the composition is described in terms of a 1D spherically symmetric profile. Such models aim to determine the spherically averaged concentration of chemical elements as a function of radius and time. In our 2D model, the analogue of a spherical average is the horizontal mean of the tracer field. Taking the horizontal mean isolates the net vertical flux, which is the relevant quantity for radial transport in stars. At the same time, there are also rapid temporal oscillations associated with IGWs. Hence, we also perform time averaging to retain only the slow-time evolution that contributes to mixing on evolutionary timescales. Thus, by combining the horizontal mean to focus on vertical transport with a time average to filter out oscillatory dynamics, we identify the corresponding net vertical transport to be used in stellar evolution models.

\section{Weakly nonlinear theory} \label{sec:result}

\subsection{Multiscale asymptotic analysis} \label{sec:theory}

To analyze the wave dynamics in the Boussinesq system \eqref{eq:Boussq}, we consider the regime of weakly nonlinear waves with nondimensional amplitude $a\ll1$. Recall that the condition for IGWs to be weakly nonlinear is $w k_z/\omega \ll 1$. In terms of dimensionless ratios, this criterion is $a \ll (k_h/k_z) \, (\omega/N)$. In this setting, we introduce a multiple-scale expansion in time 
$$t\to \tilde{t} + a^{-1} \overline{t}.$$ 
Here, the fast time scale $\tilde{t}$ resolves the rapid oscillations of the internal gravity waves, while the slow time scale $\overline{t}$ governs the gradual modulation of the wave envelope.
We define the fast-time average of a function $f \left(\mathbf{x}, \tilde{t},\overline{t} \right) $ as 
$\overline{f} \left( \mathbf{x},\overline{t} \right) =1/T \left( \int_{0}^T f \left( \mathbf{x}, \tilde{t}, \, \overline{t} \right) \, d\tilde{t} \right)$,  
where $T$ denotes the period of the fast oscillations in $\tilde{t}$. 
To isolate the vertically varying mean component from horizontal oscillations, we define the horizontal mean of a field $f \left( \mathbf{x}, \tilde{t},\overline{t} \right) $ as $\langle f\rangle \left( z,\tilde{t},\overline{t} \right)  = 1/L_x \left( \int_0^{L_x} f \left( \mathbf{x}, \tilde{t},\overline{t} \right) dx \right)$, where $L_x$ denotes the horizontal averaging length. In theoretical analyses, the average is taken over a full period of the wave field, whereas in simulations, the average is taken over the entire horizontal domain, where periodic boundary conditions are imposed in the $ x$-direction.

As the wave dynamics described by \eqref{eq:Boussq} are decoupled from the transport equation \eqref{eq:transport}, we first derive the wave solution to second order in $a$. We consider steady background fields that contain no wave component. Perturbations are then introduced up to $\mathcal{O}(a^2)$ as 
\begin{subequations} 
    \begin{align}
        \mathbf{u} \left( \mathbf{x},t \right) &= a \mathbf{u}_1  \left( \mathbf{x}, \tilde{t} \right) + a^2 \mathbf{u}_2  \left( \mathbf{x}, \tilde{t}, \overline{t} \right), \label{eq:u} \\
        b  \left( \mathbf{x},t \right) &= b_0  \left( z \right) + a b_1  \left( \mathbf{x}, \tilde{t}, \overline{t} \right) + a^2 b_2  \left(\mathbf{x}, \tilde{t}, \overline{t} \right), \\
        p  \left( \mathbf{x},t \right) &= p_0  \left( z \right) + a p_1  \left( \mathbf{x}, \tilde{t}, \overline{t} \right) + a^2 p_2  \left( \mathbf{x}, \tilde{t}, \overline{t} \right), \\
        c  \left( \mathbf{x}, t \right) &= c_0 \left(z,\bar{t} \right) + a c_1  \left( \mathbf{x}, \tilde{t}, \overline{t} \right) + a^2 c_2  \left( \mathbf{x}, \tilde{t}, \overline{t} \right).
    \end{align}
\end{subequations}
In what follows, each quantity is further decomposed into a fast-time component denoted by subscript $f$ that depends on both the fast time $\tilde{t}$ and slow time $\overline{t}$, and a slow-time component denoted by subscript $s$ dependent only on $\tilde{t}$. The asymptotic expansion is valid so long as $a^2 t\ll1$, which is sufficient considering the small amplitude waves in stars and the stochastic generation of waves due to convection.

At the leading order $\mathcal{O}(1)$, we obtain the base states
\begin{equation*}
    \nabla p_0 - b_0 \mathbf{e}_z = 0, \quad
    -\kappa \nabla^2 b_0 = 0
\end{equation*}
which admits the general solution $b_0$ linear in $z$ and $p_0$ quadratic in $z$.
Without loss of generality, we normalize the buoyancy frequency $N^2 \equiv \partial b_0/\partial z$ to unity. In stellar interiors, the background buoyancy profile $b_0(z)$ arises from stellar structure models and can exhibit complex spatial variation. However, thermal equilibrium $\nabla \cdot (\kappa \nabla b_0) = 0$ can still be maintained as the thermal diffusivity $\kappa$ itself is not constant and varies with radius $r$, allowing for significant variation in the buoyancy gradient while preserving diffusive balance.

At the next order $\mathcal{O}(a)$, the system is linear and describes the IGWs. The governing equations are 
\begin{subequations} \label{eq:1st}
    \begin{align}
        \pd{\mathbf{u}_1}{\tilde{t}} + \nabla p_1 - b_1 \mathbf{e}_z &= 0, \label{eq:u1} \\
        \pd{b_1}{\tilde{t}} - \kappa \nabla^2 b_1 + N^2 w_1 &= 0,   \label{eq:b1} \\
        \nabla \cdot \mathbf{u}_1 &= 0,
    \end{align} 
\end{subequations} 
Let the first-order wave velocity field $\mathbf{u}_1$ be periodic in $\tilde{t}$ and independent of the slow time $\overline{t}$, i.e. $\mathbf{u}_1=\mathbf{u}_{1f}(\mathbf{x},\tilde{t})$. We first take the $z-$component of the curl of the curl of \eqref{eq:u1}, yielding
\begin{equation}
    \pd{}{\tilde{t}}(-\nabla^2 w_1) + \nabla_h^2 b_1 =0. \label{eq:b1_soln} 
\end{equation}
Fast-time averaging gives $\nabla_h^2 b_{1s}\left(x,z,\bar{t}\right)=0$. Since our domain is horizontally periodic, this implies $b_{1s}$ is a function of only $z$ and $\bar{t}$, i.e. $b_{1s}=b_{1s}(z,\bar{t})$. In addition, taking the fast-time average of \eqref{eq:b1} yields $\kappa \nabla^2 b_{1s} =0$.
Applying zero boundary conditions at both top and bottom gives $b_{1s}=0$, i.e., $b_1=b_{1f}(\mathbf{x},\tilde{t})$.
Equating the fast-time component of \eqref{eq:b1} with the solution from \eqref{eq:b1_soln} leads to the linear dispersion relation.

We proceed to the next order $\mathcal{O}(a^2)$, where nonlinear interactions between first-order fields appear as source terms. The governing equations at this order are 
\begin{subequations}
    \begin{align}
        \pd{\mathbf{u}_2}{\tilde{t}} + \nabla p_2 - b_2 \mathbf{e}_z &= - \mathbf{u}_1 \cdot \nabla \mathbf{u}_1, \label{eq:u2} \\
        \pd{b_2}{\tilde{t}}  - \kappa \nabla^2 b_2 + N^2 w_2 &= - \mathbf{u}_1 \cdot \nabla b_1, \label{eq:b2} \\
        \nabla \cdot \mathbf{u}_2 &= 0 . 
    \end{align} 
\end{subequations}
We begin with the second-order momentum equation \eqref{eq:u2}. Taking the $z-$component of the curl of the curl of the momentum equation and averaging over fast time yields 
\begin{equation*}
    \nabla_h^2  \overline{b_2}  = - \overline{ \left[  \nabla \times (\nabla \times (\mathbf{u}_1 \cdot \nabla \mathbf{u}_1) ) \right]_z} 
\end{equation*}
so that 
\begin{equation}
    b_{2s} :=\overline{b_2} = - \int_0^x \int_0^{\xi} \overline{ \left[ \nabla \times (\nabla \times (\mathbf{u}_1 \cdot \nabla \mathbf{u}_1) ) \right]_z} (\xi',z,\overline{t} ) \, d\xi' \, d\xi + f (z,\overline{t}) , 
\end{equation}
where periodicity of $b_{2s}$ has been imposed. Here, the subscript $z$ denotes the $z-$component of the vector field.
The function $f \left( z,\overline{t} \right)$ arises as an integration constant and will be determined in the next step.
Taking the fast-time average of \eqref{eq:b2}, we obtain
\begin{equation} \label{eq:2nd_b1s}
    - \kappa \overline{ \nabla^2 b_2} + N^2 \overline{w_2} = - \overline{\mathbf{u}_1 \cdot \nabla b_1}. 
\end{equation}
The nonlinear term on the right-hand side $\overline{\mathbf{u}_1 \cdot \nabla b_1}$ contains both a horizontally-averaged mean component and an oscillatory component. Because the continuity equation implies $\langle w \rangle=0$, the vertical variation in $b_{2s}$ must compensate for the mean part of this forcing, i.e., we require $\kappa \partial_z^2  f \left(z,\overline{t}  \right)  = \left\langle \overline{\mathbf{u_1} \cdot \nabla b_1} \right\rangle$ in \eqref{eq:2nd_b1s}. 
Now we obtain the second-order slow-time wave velocity from \eqref{eq:2nd_b1s} which is known as the nonlinear Eulerian mean velocity 
\begin{equation} \label{eq:wE}
    w^E(x,z,\overline{t}) := \overline{w_2} = - \overline{\mathbf{u}_1 \cdot \nabla b_1} + \left\langle \overline{\mathbf{u}_1 \cdot \nabla b_1} \right\rangle + \kappa \nabla^2 \left( - \int_0^x \int_0^{\xi} \overline{ \left[ \nabla \times (\nabla \times (\mathbf{u}_1 \cdot \nabla \mathbf{u}_1) ) \right]_z} \left(\xi',z,\overline{t} \right) \, d\xi' \, d\xi \right).
\end{equation}
Note that for a single damped plane wave, the nonlinear self-advection term satisfies $\nabla \times (\nabla \times (\mathbf{u}_1 \cdot \nabla \mathbf{u}_1) )=0$, whereas the buoyancy advection term $\overline{\mathbf{u}_1 \cdot \nabla b_1}$ is nonzero. The latter represents the divergence of the wave-induced buoyancy flux. For a single wave, this term is horizontally uniform and therefore equals its horizontal average, so the first two terms in \eqref{eq:wE} cancel, in which case $w^E=0$. In contrast, for interactions between two damped plane waves, horizontally averaged buoyancy advection vanishes, i.e. $\left\langle \overline{\mathbf{u}_1 \cdot \nabla b_1} \right\rangle=0$. In this case, the remaining nonlinear terms in \eqref{eq:wE} give rise to a nonzero second-order Eulerian mean velocity $w^E(\mathbf{x},\overline{t})$.

We now turn to the transport of the passive tracer $c(\mathbf{x},\tilde{t},\overline{t})$ driven by internal gravity waves. At each order in $a$, the governing equations for the passive tracer are
\begin{subequations}
    \begin{align}
        \mathcal{O}(a): \quad \frac{\partial c_0}{\partial \overline{t}} + \frac{\partial c_1}{\partial \tilde{t}} + \frac{\partial c_0}{\partial z} w_1 &= 0, \label{eq:c1}\\
        \mathcal{O}(a^2): \quad \pd{c_1}{\overline{t}} + \pd{c_2}{\tilde{t}} + \frac{\partial c_0}{\partial z} w_2 &= -\mathbf{u}_1 \cdot \nabla c_1. \label{eq:c2}
    \end{align}
\end{subequations}
Taking the fast-time average of \eqref{eq:c1} gives
$\partial c_0/\partial \overline{t}=0$, indicating that the leading-order tracer field $c_0=c_0(z)$ is time-independent. The fast-time component $c_{1f}(\mathbf{x},\tilde{t})$ can then be determined from \eqref{eq:c1} as 
\begin{equation}
    c_{1f}(\mathbf{x},\tilde{t}) = - \frac{dc_0}{dz} \int_0^{\tilde{t}} w_1(\mathbf{x},\tau) \, d\tau, \label{eq:c1f}
\end{equation}
while the slow component $c_{1s}:=\overline{c_1}$ enters at the next order.
Taking the fast-time average of \eqref{eq:c2} yields the net transport in $c_1$ as 
\begin{equation}
    \pd{c_{1s}}{\overline{t}} := \pd{c_1}{\overline{t}}= - \overline{\mathbf{u}_1 \cdot \nabla c_{1f}} - \frac{d c_0}{dz} \overline{w_2} = -\mathbf{u}^S \cdot \nabla c_0 - \frac{dc_0}{dz} \overline{w}_2 = -\frac{dc_0}{dz} (w^S+w^E), \label{eq:c1s}
\end{equation}
where
\begin{equation}
    \mathbf{u}_1^S = \overline{(\boldsymbol{\xi} \cdot \nabla) \mathbf{u}_1}, 
\end{equation}
denotes the Stokes drift with the displacement $\boldsymbol{\xi}$ of fluid parcels given by $\partial \boldsymbol{\xi}/\partial \tilde{t}=\mathbf{u}_1$, $\mathbf{u}^E:=\overline{\mathbf{u}}_2$ represents the Eulerian mean velocity, and the identity $\overline{\mathbf{u}_1 \cdot \nabla c_{1f}} =\mathbf{u}^S \cdot \nabla c_0 $ follows from integration by parts.


As shown below, the horizontal mean of $c_{1s}$ vanishes, so the leading-order tracer response does not produce a net vertical transport. A nonzero horizontally averaged transport therefore arises at higher order. Although the passive-tracer equation could be expanded directly to $\mathcal{O}(a^3)$, this introduces many intermediate terms that ultimately cancel.
Instead, it is more convenient to first derive an exact relation for the horizontally averaged tracer transport and then apply the asymptotic expansion.

We decompose the tracer and velocity into horizontal means and fluctuations $c=\langle c\rangle + c'$, and $\mathbf{u}=\langle \mathbf{u}\rangle + \mathbf{u}'$.
Using incompressibility and horizontal periodicity, the horizontal average of the transport equation \eqref{eq:transport} gives
\begin{equation} \label{eqn:net transport}
    \partial_t \langle c\rangle + \partial_z\langle w c \rangle = 0.
\end{equation}
An equation for the tracer variance is obtained by multiplying the fluctuation equation by $c'$ and taking the horizontal average
\begin{equation}
    \partial_t \left\langle\frac{c'^2}{2}\right\rangle + \left\langle\mathbf{u}\cdot\nabla \frac{c'^2}{2}\right\rangle + \langle w' c' \rangle \partial_z\langle c\rangle = 0. 
\end{equation}
Since $\langle w \rangle=0$, this equation provides an expression for the net vertical flux of trace, $\langle w' c' \rangle$.
Eliminating $\langle w'c'\rangle$ between the mean and variance equations therefore gives the exact relation
\begin{equation}
\partial_t\langle c \rangle = \partial_z\left[\frac{1}{2}
    \left(\partial_z \langle c \rangle \right)^{-1}\left(\partial_t\left\langle c'^2\right\rangle + \partial_z\left\langle w' c'^2\right\rangle\right) \right].
\end{equation}
We now substitute the asymptotic expansions into this relation and average over the fast time. At the first two orders, $\partial_{\bar{t}} \langle c_1 \rangle = \partial_{\tilde{t}} \langle c_1 \rangle =0$. Since the initial tracer profile contains no first-order perturbation, it follows that $\langle c_1 \rangle=0$. The first nonzero contribution to the horizontally averaged slow tracer field is therefore $\langle c_{2s}\rangle$. At $\mathcal{O}(a^3)$, the averaged equation becomes
\begin{equation} \label{eq:c2s_old}
    \partial_{\bar{t}}\langle c_{2s} \rangle = \partial_z\left(\frac{1}{2}
    \left(\frac{dc_0}{dz}\right)^{-1}\partial_{\bar{t}}\overline{\langle c_1^2\rangle}\right),
\end{equation}
where we have used that $\overline{\langle w_1 c_1^2\rangle}=0$ (see Appendix \ref{append:proof}).
The slow-time dependence of $\overline{\langle c_1^2\rangle}$ arises from $c_{1s}^2$. Using \eqref{eq:c1s}, we obtain
\begin{equation}
    \label{eq:c2s}
    \partial_{\bar{t}}\langle c_{2s} \rangle=\partial_z \left(\bar{t}\langle(w^S+w^E)^2\rangle
    \frac{dc_0}{dz}\right).
\end{equation}

Equation \eqref{eq:c2s} is the main result for the leading-order net vertical transport of a passive tracer by a coherent superposition of internal gravity waves. It has the form of a diffusion equation, with a time-dependent effective diffusivity $\bar{t}\langle(w^S+w^E)^2\rangle$. 
In the following, we evaluate $\langle(w^S+w^E)^2\rangle$ for superpositions of propagating internal gravity waves.

\subsection{Wave interaction decomposition} \label{sec:result_interact}

We consider a superposition of waves $j$ with first-order vertical velocity fields given by
\begin{align} \label{eq:w1}
    w_1(x,z,\tilde{t}) &= \sum_{j=1}^{n} A_j  e^{-z/\mu_j} \cos \left( k_{h,j} x + k_{z,j} z - \omega_j \tilde{t} + \varphi_j \right), 
\end{align}
where $A_j >0$ and $A_j=\mathcal{O}(1)$ is the relative amplitudes of the different wave components in the superposition, $k_{h,j}\in\mathbb{R}$ is the horizontal wavenumber, $\omega_j>0$ is the frequency, $k_{z,j}<0$ is the vertical wavenumber corresponding to upward propagation, and $\varphi_j$ is the phase. Both the vertical wavenumber $k_{z,j}=k_{z,j}(k_{h,j},\omega_j)$ and the damping rate $\mu_j=\mu_j(k_{h,j},\omega_j)$ are determined by the linear dispersion relation.

Under the initial condition $c_0=c_0(z)$, a single vertically damped weakly nonlinear internal gravity wave produces only oscillatory tracer perturbations with zero time-averaged tracer transport.
Therefore, we must consider the superposition of multiple waves to permit nonlinear resonant interactions between modes with different wavenumbers or frequencies \citep{Jermyn2022}.

In the first-order net transport equation \eqref{eq:c1s}, the evolution of $c_{1s}$ is driven by the interaction of wave pairs. These interactions manifest through two contributions. The first, the Stokes drift, is the correlation between the first-order wave velocity field and the first-order passive tracer perturbation $u_1 \cdot \nabla c_{1f}$. The second is the advection of the background tracer field $c_0$ by the nonlinear Eulerian mean velocity $w^E$, which itself is due to nonlinear interactions between a pair of waves. In the absence of radiative diffusion, the Eulerian mean vertical velocity exactly balances the Stokes drift, i.e. $w^E=-w^S$, resulting in $c_{1s}=0$ and therefore no net transport. This cancellation reflects the conservation of pseudomomentum \citep[e.g.,][]{andrews1978wave, scinocca1992nonlinear, shaw2008wave}. However, when the waves experience any type of dissipation, the Stokes drift and Eulerian mean flow do not cancel. If we include the effects of radiative diffusion, we obtain 
\begin{equation} \label{eq:c1s_int}
        \begin{split}
                c_{1s}(x,z,\overline{t}) = \overline{t} \, \pd{c_0(z)}{z}  
                \sum_{i=1}^n \sum_{\substack{j=1 \\ j\neq i }}^n  e^{-z\left(\frac{1}{\mu_i}+\frac{1}{\mu_j}\right)}\delta_{\omega_i,\omega_j} \Big[ &C_1^{(i,j)} \cos \left( (k_{h,i}-k_{h,j})x + (k_{z,i}-k_{z,j})z + \varphi_i-\varphi_j \right) \\
                &+ C_2^{(i,j)} \sin \left( (k_{h,i}-k_{h,j})x + (k_{z,i}-k_{z,j})z + \varphi_i-\varphi_j \right) \Big], 
        \end{split}
\end{equation}
where $C_{1,2}^{(i,j)}(A_{\{i,j\}},k_{h,\{i,j\}},k_{z,\{i,j\}},\omega_{\{i,j\}},\mu_{\{i,j\}},\kappa)$ are coefficients determined by the wave parameters and the radiative diffusivity. The subscripts $i$ and $j$ refer to the two distinct interacting wave components. For more than two waves, we sum over all pairs of two waves. If $\omega_i\neq\omega_j$, the fast-time average of the terms on the right-hand side of \eqref{eq:c1s_int} is zero. This implies that the net tracer deviation only occurs when $\omega_i=\omega_j$ (but $k_{h,i}\neq k_{h,j}$), yielding a contribution that scales as $\overline{c}-c_0=\mathcal{O}(a^2t)$. For more than two waves, we sum over all pairs of waves with equal frequency. However, at this order, the horizontal average of the transport vanishes, resulting in zero net vertical transport.

For net vertical transport, we use equation \eqref{eq:c2s}.
Considering the weakly nonlinear velocity field \eqref{eq:w1}, we obtain
\begin{subequations}  
    \begin{align} \label{eq:c2s_n}
    \left\langle c_{2s} (z,\bar{t}) \right\rangle &= 
    \frac{\bar{t}^{2}}{2} \partial_z  \left[ \frac{dc_0}{dz} \sum_{i=1}^{n} \sum_{j=1}^{n} \sum_{k=1}^{n} \sum_{l=1}^{n} 
    \delta_{\omega_i,\omega_j} \, \delta_{\omega_k,\omega_l}
    ~ e^{-z\left(\frac{1}{\mu_i}+\frac{1}{\mu_j}+\frac{1}{\mu_k}+\frac{1}{\mu_l}\right)} A_i A_j A_k A_l  \,
    \big\langle
    \alpha_1  \cos (\Delta)
    + \alpha_2 \sin (\Delta)
    \big\rangle  \right], 
\end{align}
with phase
\begin{align} \label{eq:c2s_phase}
    \Delta &= \left(k_{h,i}-k_{h,j}+k_{h,k}-k_{h,l}\right)x + \left(k_{z,i}-k_{z,j}+k_{z,k}-k_{z,l}\right)z + \varphi_i-\varphi_j+\varphi_k-\varphi_l. 
\end{align}
\end{subequations}
The coefficients $\alpha_m=\alpha_m(k_{h,\{i,j,k,l\}},k_{z,\{i,j,k,l\}},\mu_{\{i,j,k,l\}},\omega_{\{i,j,k,l\}},\kappa,N^2)$, $m=1,2$, are constants determined by the wave parameters and background stratification (see the derivation in Appendix \ref{append:net}). 
Nonzero vertical transport arises when the frequency resonance condition is satisfied and the horizontal wavenumbers $(k_{h,i},k_{h,j},k_{h,k},k_{h,l})$ simultaneously satisfy the resonant conditions in \eqref{eq:c2s_phase}.
This yields
\begin{equation}
    \langle \overline{c}\rangle-c_0 = a^2 \langle c_{2s} \rangle =\mathcal{O}(a^4 t^2).
\end{equation}
In the special case of an interaction between two waves $i$ and $j$, the evolution equation reduces to
\begin{equation}
     \langle\overline{c}\rangle -c_0 = a^4  D_c \partial_z \left( e^{-z/(\frac{1}{\mu_i}+\frac{1}{\mu_j})} \partial_z c_0(z) \right) t^2,
\end{equation}
where $D_c(A_{\{i,j\}},k_{h,\{i,j\}},k_{z,\{i,j\}},\mu_{\{i,j\}},\omega_{\{i,j\}},\kappa,N^2)$ is depends on the wave parameters. Note that the net transport $\langle \overline{c}(z,t) \rangle$, or equivalently $\langle c_{2s} \rangle (z,t)$, exhibits no oscillation in $z$ for 2--wave interactions. This is because, whenever the horizontal wavenumbers $(k_{h,i},k_{h,j},k_{h,k},k_{h,l})$ in \eqref{eq:c2s_phase} \eqref{eq:c2s_phase} satisfy the resonance conditions for a nonzero horizontal mean, the corresponding vertical wavenumbers $(k_{z,i},k_{z,j},k_{z,k},k_{z,l})$ simultaneously satisfy the resonance conditions as well.

\section{Simulation results} \label{sec:sims}

We examine transport by $2-$, $3-$, and $4-$wave nonlinear interactions using the simulation method described in Section \ref{sec:sim_setup}. The results from the two-dimensional simulations are compared with the asymptotic results presented in Section \ref{sec:result_interact}.
The theoretical prediction indicates that net transport driven by nonlinear wave interactions scales as $\overline{c}-c_0 = \mathcal{O}(a^2t)$ and $\langle \overline{c} \rangle - c_0 =  \mathcal{O}(a^4 t^2)$. Given the small magnitude of transport at low wave amplitudes, we employ wave amplitude $a\sim10^{-3}$ and radiative diffusivity $\kappa=5\cdot10^{-4}$ in the simulations so that the resulting transport can be captured within numerical precision. However, these parameters are carefully selected to avoid artificial wave reflection at the top boundary while preserving physical plausibility. In the following, we fix the initial passive tracer field $c_0:=c(z,t=0)=z^2$. The phases of the wave components in \eqref{eq:w1} are set to $\varphi_j=0$.

We compare the simulations to the predictions of the asymptotic calculation in two ways. In our simulations, we slowly turn on the wave forcing at $t_0$ and slowly turn off the wave forcing at $t_1$. The first prediction is that, when the wave amplitudes are roughly constant between the times, we expect $\overline{c}(x_0, z_0,t)-c_0(z_0)$ to increase linearly with time, and for $\langle\overline{c}\rangle(z_0,t) - c_0(z_0)$ to increase quadratically in time.
Specifically, we write 
\begin{align} \label{eq:sim_fit}
    \overline{c}(x_0,z_0,t)-c_0 = \beta (t-t_{\rm fit}), \quad 
    \langle \overline{c}\rangle(z_0,t)-c_0 = \gamma_1(t-t_{\rm fit})^2+\gamma_2,
\end{align}
where $\beta$ and $\gamma_1$ are the transport coefficients, while $t_{\rm fit}$ and $\gamma_2$ are fitting parameters introduced in the simulations to account for wave initialization and shut-down effects. 
Note that $t_{\rm fit}$ cannot be determined from the intercept of the linear relation and is therefore obtained from the quadratic fit.
Table \ref{tab} lists the transport coefficients $\beta$ and $\gamma_1$ obtained from simulations, as well as their theoretical values calculated via asymptotics given by \eqref{eq:c1s_int} and \eqref{eq:c2s_n}.

The second prediction is the total change in $c$ after the waves have turned off at times $t > t_1$.
We examine $\overline{c}(x_0,z)-c_0$ and $\langle \overline{c}\rangle(z)-c_0$ as functions of $z$.
Since the turning-on and turning-off process in simulations introduces transient effects, the simulation profiles at the final time are fitted to \eqref{eq:c1s_int} and \eqref{eq:c2s_n} using $t_{\rm transport}$ as a fitting parameter for $\bar{t}$. 
The fitted $t_{\rm transport}$ is then compared with the estimated transport duration in the simulation $\Delta t \approx t_1-t_0=1050\pi\approx3300$. The fitted values of $t_{\rm transport}$ for the test cases are given in Table \ref{tab}.

\begin{table}[tb!]
\hspace{-1in}
\begin{tabular}{|c|c|ccc|ccccc|}
\hline
\multirow{3}{*}{Case} & \multirow{3}{*}{Fig.} & \multicolumn{3}{c|}{$c(x_0,z,t)-c_0$} & \multicolumn{5}{c|}{$\langle \overline{c}\rangle(z,t)-c_0$} \\ \cline{3-10} 
&  & \multicolumn{1}{c|}{theory} & \multicolumn{2}{c|}{sim} & \multicolumn{1}{c|}{theory} & \multicolumn{4}{c|}{sim} \\ \cline{3-10} 
&  & \multicolumn{1}{c|}{$\beta$} & \multicolumn{1}{c|}{$\beta$} & $t_{\rm transport}$ & \multicolumn{1}{c|}{$\gamma_1$} & \multicolumn{1}{c|}{$\gamma_1$} & \multicolumn{1}{c|}{$\gamma_2$} & \multicolumn{1}{c|}{$t_{\rm fit}$} & $t_{\rm transport}$ \\ \hline
Two waves & 1,2,3 & \multicolumn{1}{c|}{$-1.34\cdot10^{-6}$} & \multicolumn{1}{c|}{$-1.34\cdot10^{-6}$} & $3173$ & \multicolumn{1}{c|}{$-1.72\cdot10^{-13}$} & \multicolumn{1}{c|}{$-1.73\cdot 10^{-13}$} & \multicolumn{1}{c|}{$-3.59\cdot10^{-6}$} & \multicolumn{1}{c|}{$604$} &  $3186$\\ \hline
Three waves & 4 & \multicolumn{1}{c|}{$-1.31\cdot10^{-6}$} & \multicolumn{1}{c|}{$-1.31\cdot10^{-6}$} & $3168$ & \multicolumn{1}{c|}{$-1.68\cdot10^{-13}$} & \multicolumn{1}{c|}{$-1.62\cdot10^{-13}$} & \multicolumn{1}{c|}{$3.92\cdot10^{-6}$} & \multicolumn{1}{c|}{543} & 3187 \\ \hline
Four waves case (a) & 5 & \multicolumn{1}{c|}{$-1.32\cdot10^{-7}$} & \multicolumn{1}{c|}{$-1.15\cdot10^{-7}$} & 3084 & \multicolumn{1}{c|}{$-3.29\cdot10^{-13}$} & \multicolumn{1}{c|}{$-3.31\cdot10^{-13}$} & \multicolumn{1}{c|}{$-8.46\cdot10^{-6}$} & \multicolumn{1}{c|}{657} & 3182 \\ \hline
Four waves case (c,1) & 6 & \multicolumn{1}{c|}{$4.61\cdot10^{-6}$} & \multicolumn{1}{c|}{$4.69\cdot10^{-6}$} & 3207 & \multicolumn{1}{c|}{$-1.49\cdot10^{-13}$} & \multicolumn{1}{c|}{$-1.56\cdot10^{-13}$} & \multicolumn{1}{c|}{$5.08\cdot10^{-6}$} & \multicolumn{1}{c|}{542} & 3228 \\ \hline
Four waves case (c,2) & 7 & \multicolumn{1}{c|}{$-1.47\cdot10^{-6}$} & \multicolumn{1}{c|}{$-1.46\cdot10^{-6}$} & 3154 & \multicolumn{1}{c|}{$-2.12\cdot10^{-13}$} & \multicolumn{1}{c|}{$-1.87\cdot10^{-13}$} & \multicolumn{1}{c|}{$2.55\cdot10^{-5}$} & \multicolumn{1}{c|}{1108} & 3157 \\ \hline
\end{tabular}
\caption{Comparison of theoretical predictions and simulation results for net transport coefficients. Columns list the transport coefficient $\beta$ for $c(x_0,z,t)-c_0$, the quadratic coefficient $\gamma_1$ for vertical transport $\langle \overline{c}\rangle(z,t)-c_0$, the fitted parameters $\gamma_2$, $t_{\rm fit}$ in simulation using \eqref{eq:sim_fit}, and the fitted effective transport time $t_{\rm transport}$. Results are shown for the 2--wave, 3--wave, and two representative 4--wave cases.}
\label{tab}
\end{table}

\subsection{Two waves}

\begin{figure}[tb!]
    \centering
    \includegraphics[width=\linewidth]{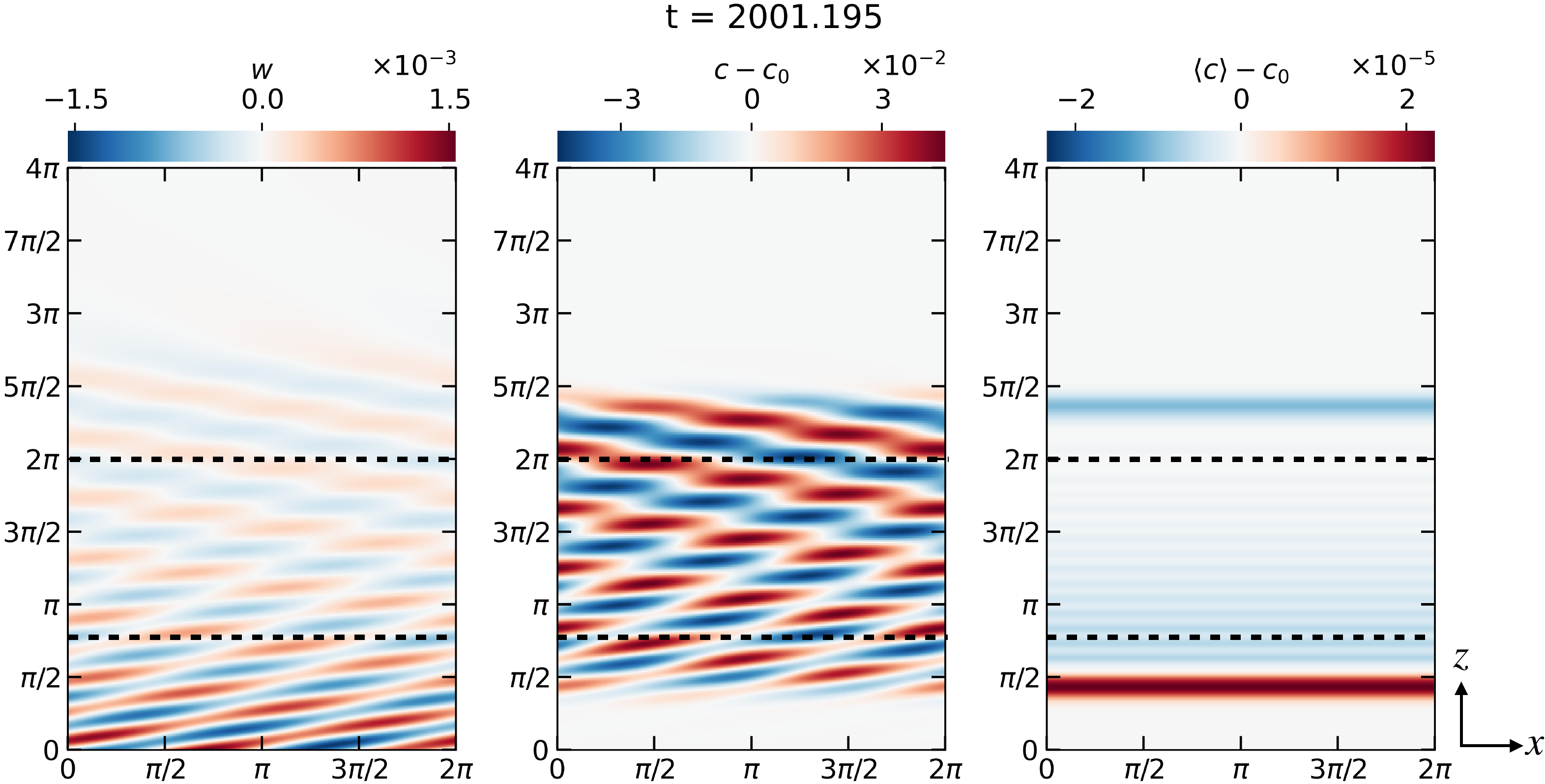}
    \caption{Snapshots of a simulation of two waves during the wave forcing, showing the vertical velocity $w$, passive tracer $c(x,z)$, and the horizontal mean of the passive tracer $\langle c(z) \rangle$. Our subsequent analysis is confined to the region $(x,z)\in[0,2\pi]\times[0.8\pi,2\pi]$ delineated by the black dashed lines to avoid boundary artifacts.}
    \label{fig:snapshot}
\end{figure}

Figure \ref{fig:snapshot} shows the transport induced by two sinusoidal waves with parameters \begin{equation} \label{eq:2waves_param}
    (a_1,k_{h,1},\omega)=(0.002,2,0.2), \quad (a_2,k_{h,2},\omega)=(0.0005,-1,0.2)
\end{equation}
forced at the boundary $z=0$. 
The first column of Fig.~\ref{fig:snapshot} displays the vertical velocity field at an intermediate time. Due to radiative damping, the wave amplitude decays with $z$. The multiple wave components produce spatially oscillatory patterns while carrying energy upward with positive vertical group velocity.
The second column in Fig.~\ref{fig:snapshot} shows the change in the passive tracer induced by the waves. 
The third column displays the horizontally averaged passive tracer field $\langle c(z,t) \rangle$ representing the vertical transport. 
There are strong positive and negative transport signals near $z=\pi/2$ and $z=5\pi/2$, corresponding to the transition regions where the nonlinear advection of $c$ is switched on. These features are artifacts of the numerical treatment of the tracer equation and are therefore excluded from the analysis by restricting attention to the region between the dashed horizontal lines.

\begin{figure}[tb!]
    \centering
    \includegraphics[width=\linewidth]{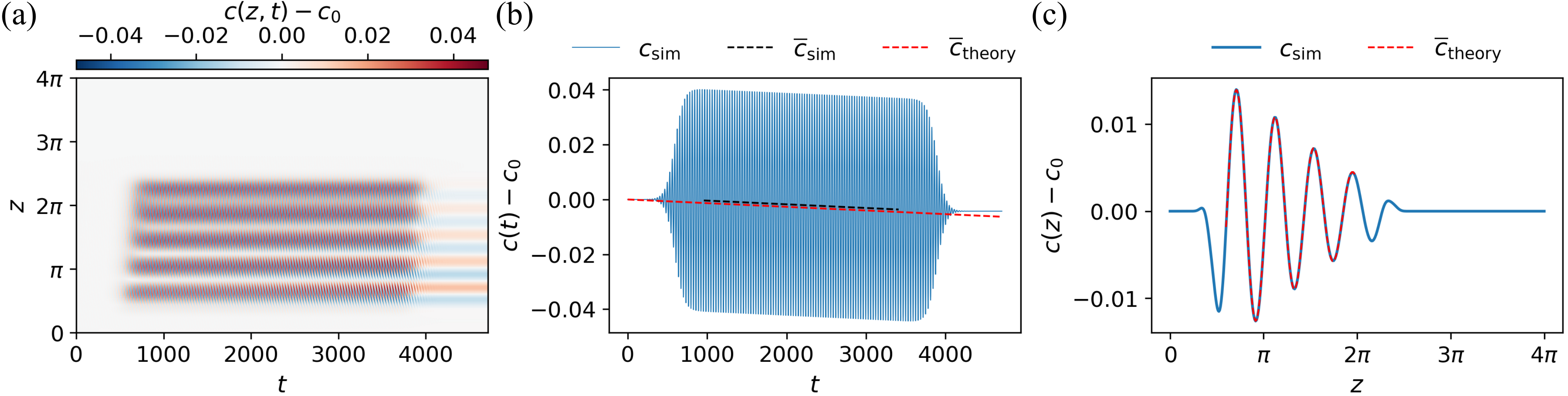}
    \caption{Transport $c(x,z,t)-c_0$ driven by two waves at (a) $x_0=\pi$, (b) $x_0=\pi$, $z_0=0.998\pi$, and (c) $x_0=\pi$, $t=4500$. 
    The wave parameters are given in \eqref{eq:2waves_param}.
    The black curve in (b) shows the period-averaged transport over the interval  $t=30~T$ to $t=115~T$, where $T$ is the wave period. Fitting the curve with the linear function in \eqref{eq:sim_fit} gives the coefficient $\beta$.
    The red curve in (b) corresponds to the theoretical prediction from \eqref{eq:c1s_int} at $x_0=\pi$, $z_0=0.998\pi$. In (c), the red curve shows theory \eqref{eq:c1s_int} at $x_0=\pi$, evaluated at $\bar{t}=t_{\rm transport}$, where $t_{\rm transport}$ is the fitted effective transport time. 
    The fitted and theoretical parameters are reported in Table \ref{tab}. }
    \label{fig:c_2waves}
\end{figure}

In Fig.~\ref{fig:c_2waves}, we extract the passive tracer $c-c_0$ from the 2--wave interaction simulation shown in Fig.~\ref{fig:snapshot} at $x_0=\pi$. 
Figure \ref{fig:c_2waves}(a) shows that $c=c_0$ from $t=0$ to $t\approx 200$ during the wave turn-on period. Also, since the tracer advection equation is solved only over $z\in[0.8\pi,2\pi]$, the tracer remains at its initial value $c=c_0$ outside this region.
After $t\approx4000$ when the waves are turned off, $c-c_0$ is nonzero indicating that net transport has occurred at $x_0=\pi$. 
In Fig.~\ref{fig:c_2waves}(b), we also fix $z_0=0.998\pi$. The simulated tracer field $c(t)-c_0$ oscillates due to local wave-driven transport. 
To quantify long-time transport, we wish to perform a fast-time average to uncover the slow-time variation. Specifically, the tracer field was first interpolated in time using cubic interpolation onto a fine grid with resolution $0.001~T$, where $T=2\pi/\omega$. Within the window $t\in[30,115]~T$, we computed period-averaged means of $c(t)-c_0$ by integrating over each wave period $T$. The temporal evolution of this fast-time averaged quantity is shown in black. Table~\ref{tab} reports the slope of a linear fit to this curve, measuring the transport at this location. 
The red curve represents the second-order asymptotic prediction from equation \eqref{eq:c1s_int} evaluated with the wave parameters given in \eqref{eq:2waves_param}.
The theoretical prediction agrees well with the mean change in $c(t)$, which captures the slow-time evolution of the passive tracer.
Note that the vertical shift between the red (theory) and black (time-averaged simulation) dashed lines is due to the wave turn-on effect, as the simulated transport does not begin at $t=0$ and $t_{\rm fit}$ is not obtained by linear fitting. The black and red curves have the same slope, showing agreement between the simulation and theory.

Figure \ref{fig:c_2waves}(c) shows $c(z)-c_0$ at $x_0=\pi$ and final time $t=4500$, illustrating the net transport resulting from the two waves. We focus on the region $z\in[0.8\pi,2\pi]$, because near the bottom ($z\lessapprox0.8\pi$) we do not include advection of $c$, and near the top ($z\gtrapprox2\pi$) the waves enter a damping layer. The simulation is compared with the theoretical prediction from equation \eqref{eq:c1s_int} at $x_0=\pi$ with a fitted evolution time $t=t_{\rm transport}$ given in Table \ref{tab}. This fitted time represents the effective duration of transport, excluding the transient intervals when the wave forcing ramps on and off. The theoretical profile again shows good agreement with the simulation, and the fitted $t_{\rm transport}$ is qualitatively consistent with the expected wave transport timescale $t_1-t_0=1050\pi$.

\begin{figure}[tb!]
    \centering
    \includegraphics[width=\linewidth]{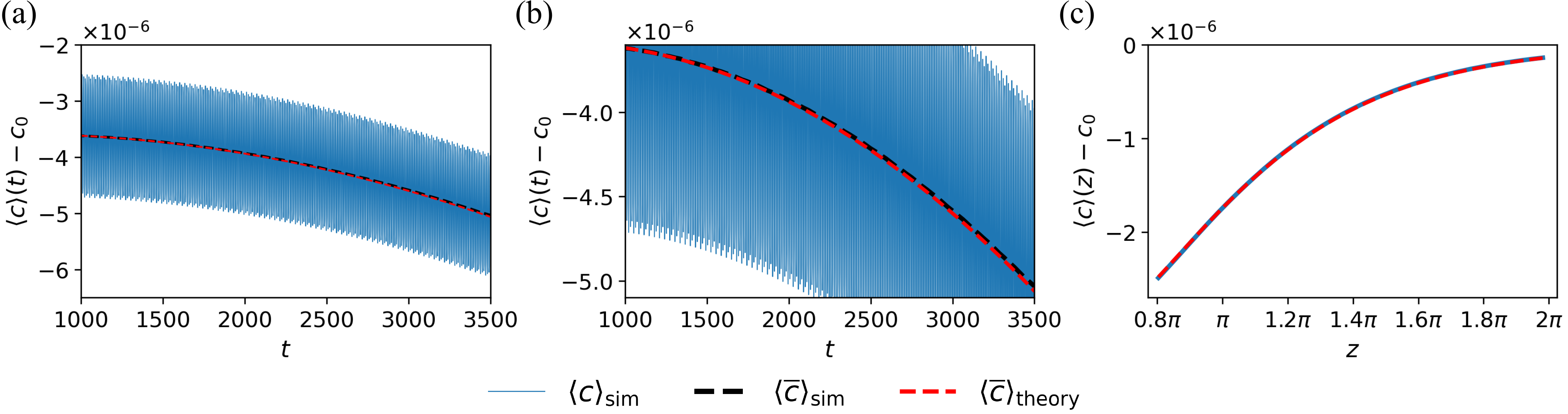}
    \caption{Vertical transport $\langle c\rangle(z,t)-c_0$ induced by two waves. The blue curves show the simulation results for the horizontally averaged tracer $\langle c \rangle(z,t)$ at (a,b) $z_0=0.998\pi$ and (c) $t=4500$. Panel (b) shows a zoomed-in view of (a). The black dashed curves in (a,b) show the period-averaged vertical transport, which is fitted with the quadratic function in \eqref{eq:sim_fit} to determine $(\gamma_1,\gamma_2,t_{\rm fit})$. The red curves show the corresponding theoretical predictions at $z_0$. 
    In (c), the simulation is compared with the theoretical prediction \eqref{eq:c2s_n} evaluated at the fitted effective transport time $\bar{t}=t_{\rm transport}$.
    A comparison between the fitted parameters and the theoretical values is provided in Table \ref{tab}. }
    \label{fig:cv_2waves}
\end{figure}

Next, we examine the net vertical transport in the simulation shown in Fig.~\ref{fig:snapshot} by analyzing the horizontal mean $\langle c \rangle(z,t)$. To avoid the influence of transient effects associated with wave initiation and termination, as well as boundary-related artifacts, we first restrict our analysis to the region $t\in[1500,3500]$ and $z\in[0.8\pi,2\pi]$. 
Figures \ref{fig:cv_2waves}(a) and (b) show the evolution of $\langle c \rangle(t)-c_0$ at fixed $z_0=0.998\pi$ over time. We apply the same time-averaging method described earlier to extract the net vertical transport, and compare the result with the theoretical prediction from \eqref{eq:c2s_n} at the same $z_0$. For the red (theory) curve, we introduce a horizontal offset $t_{\rm fit}$ to take into account the slow ramping up of the wave forcing. Nevertheless, the coefficient governing the slow-time transport shows good agreement with the simulation.
In Fig.~\ref{fig:cv_2waves}(c), we examine $\langle c \rangle(z)-c_0$  at $t=4500$ to evaluate the net vertical transport after the waves have been turned off and no further transport occurs. In this case, we compare the simulation with the theoretical prediction from \eqref{eq:c2s_n} using fitted time $t=t_{\rm transport}$ given in Table \ref{tab} to represent the effective duration of transport. 
The comparison shows that the slow-time theoretical predictions for both $c(x_0,z,t)$ and $\langle c\rangle(z,t)$ given in Section \ref{sec:result_interact} agree well with the simulation results for two waves.

\subsection{Three waves} \label{sec:sim_3wave}

\begin{figure}[tb!]
    \centering
    \includegraphics[width=\linewidth]{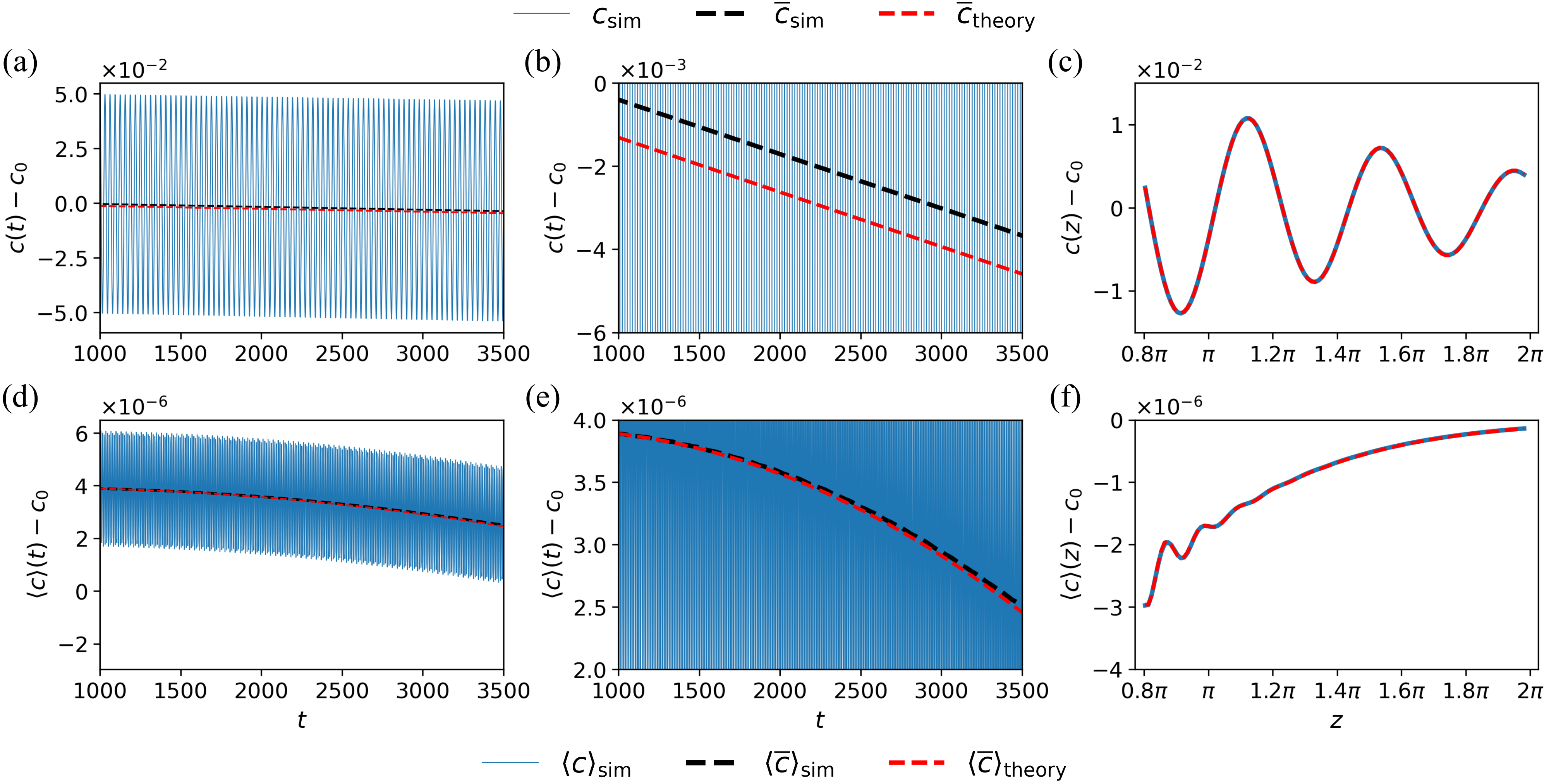}
    \caption{Transport induced by three waves with identical frequencies given in \eqref{eq:3wave_param}. Blue curves show simulation results of (a,b) $c(t)-c_0$ at $x_0=\pi$ and $z_0=0.998\pi$, (c) $c(t)-c_0$ at $x_0=\pi$ and $t=4500$, (d,e) $\langle c \rangle(z,t)-c_0$ at $z_0=0.998\pi$, and (f) $\langle c \rangle(z,t)-c_0$ at $t=4500$. The middle panels (b,e) are zoomed-in views of the left (a,d). The black curves in (a,b,d,e) are the time-averaged transport evolution. The red curves show the theoretical predictions from \eqref{eq:c1s_int} and \eqref{eq:c2s_n} at the corresponding fixed spatial location. In (c,f), the theoretical curves are evaluated at the fitted effective transport time $\bar{t}=t_{\rm transport}$.
    A comparison between the fitted parameters and theoretical values is provided in Table \ref{tab}. }
    \label{fig:c_3waves}
\end{figure}

We consider the following three cases of transport driven by three waves characterized by wave parameters $(a_{\{1,2,3\}}, k_{h,\{1,2,3\}}, \omega_{\{1,2,3\}})$ forced at the boundary $z=0$. Note that the ordering of the wave indices $1,2,3$ is interchangeable.  
\begin{enumerate}[label=(\alph*)]
    \item $\omega_1=\omega_2=\omega_3$. In this case, wave transport is expected in both $\overline{c}(x,z,t)$ and $\langle \overline{c}(z,t) \rangle$. We choose the following wave parameters for the simulation given by $\omega_j =0.2$, $j=1,2,3$, and 
    \begin{equation} \label{eq:3wave_param}
    (a_1,k_{h,1})=(0.001,-2), \quad (a_2,k_{h,2})=(0.001,1), \quad (a_3,k_{h,3})=(0.005,4).
    \end{equation}
    In Figs.~\ref{fig:c_3waves}(a) and (b), the oscillations of $c(t)-c_0$ at fixed $x_0=\pi$ and $z_0=0.998\pi$ show the local transport.
    As before, after a fast-time average, we find $c(x_0, z_0, t)$ decreases linearly with time.
    We then compare this resulting transport from the simulation with the theoretical prediction in Section \ref{sec:result_interact} by summing over the linear solutions $c_{1s}(x,z,\overline{t})$ derived in equation \eqref{eq:c1s_int}, evaluated for each pairwise combination of the three wave parameters. 
    The simulated transport $c(z)-c_0$ in Fig.~\ref{fig:c_3waves}(c) extracted at $x_0=\pi$ and at the final time illustrates the overall change in the passive tracer driven by the three waves. This result is compared with the theoretical prediction $c_{1s}(x,z,\overline{t})$ using a fitted effective transport time $t_{\rm transport}$.

    The net vertical transport $\langle c \rangle(z,t)-c_0$ is shown in Figs.~\ref{fig:c_3waves}(d,e) and (f), plotted as a function of time at fixed $z$ and as a function of $z$ at the final time $t$, respectively. The theoretical prediction is computed from equation \eqref{eq:c2s_n}, summing the permutations of the wave parameters given in \eqref{eq:3wave_param}. Good agreements between simulation and theory validate the theoretical framework for transport by three waves. Note that for the parameter choices in this simulation, both 2--wave interactions, where the indices $i,j,k,l$ in \eqref{eq:c2s_n} are chosen among only two of the waves, as well as 3--wave interactions, where the indices are chosen among all three waves, are present. The oscillatory dependence on $z$ shown in the net vertical transport $\langle c \rangle (z)-c_0$ in Fig.~\ref{fig:c_3waves}(f) arises from the 3--wave interaction.

    \item $\omega_1=\omega_2\neq\omega_3$. The 3--wave transport in both $\overline{c}(x,z,t)$ and $\langle \overline{c}(z,t) \rangle$ is identical to the 2--wave transport driven by waves 1 and 2 alone. This is because the leading-order net transport term $c_{1s}(x,z,\overline{t})$ in \eqref{eq:c1s_int} arises solely from the interaction between waves 1 and 2, which share the same frequency. Since $\omega_3$ differs, it cannot participate in the frequency-resonant interaction that generates $c_{1s}(x,z,\overline{t})$.
    For the vertical transport $c_{2s}(z,\overline{t})$,
    all the fourth-order interactions are built upon the leading-order net transport $c_{1s}(x,z,\overline{t})$, and thus wave 3 does not contribute.

    \item $\omega_1\neq\omega_2\neq\omega_3$. The 3--wave interaction does not generate net local transport since $c_{1s}(x,z,\overline{t})=0$ in \eqref{eq:c1s_int}, and no fourth-order interaction can occur in the case of a zero $\overline{c}_{1s}$. Therefore, we obtain $\overline{c}(x,z,t)=0$ and $\langle \overline{c}(z,t) \rangle=0$. 

\end{enumerate}

\subsection{Four waves}

Consider the following five cases of four waves with wave parameters $(a_{\{1,2,3,4\}}, k_{h,\{1,2,3,4\}}, \omega_{\{1,2,3,4\}})$ at the boundary $z=0$.

\begin{figure}[tb!]
    \centering
    \includegraphics[width=\linewidth]{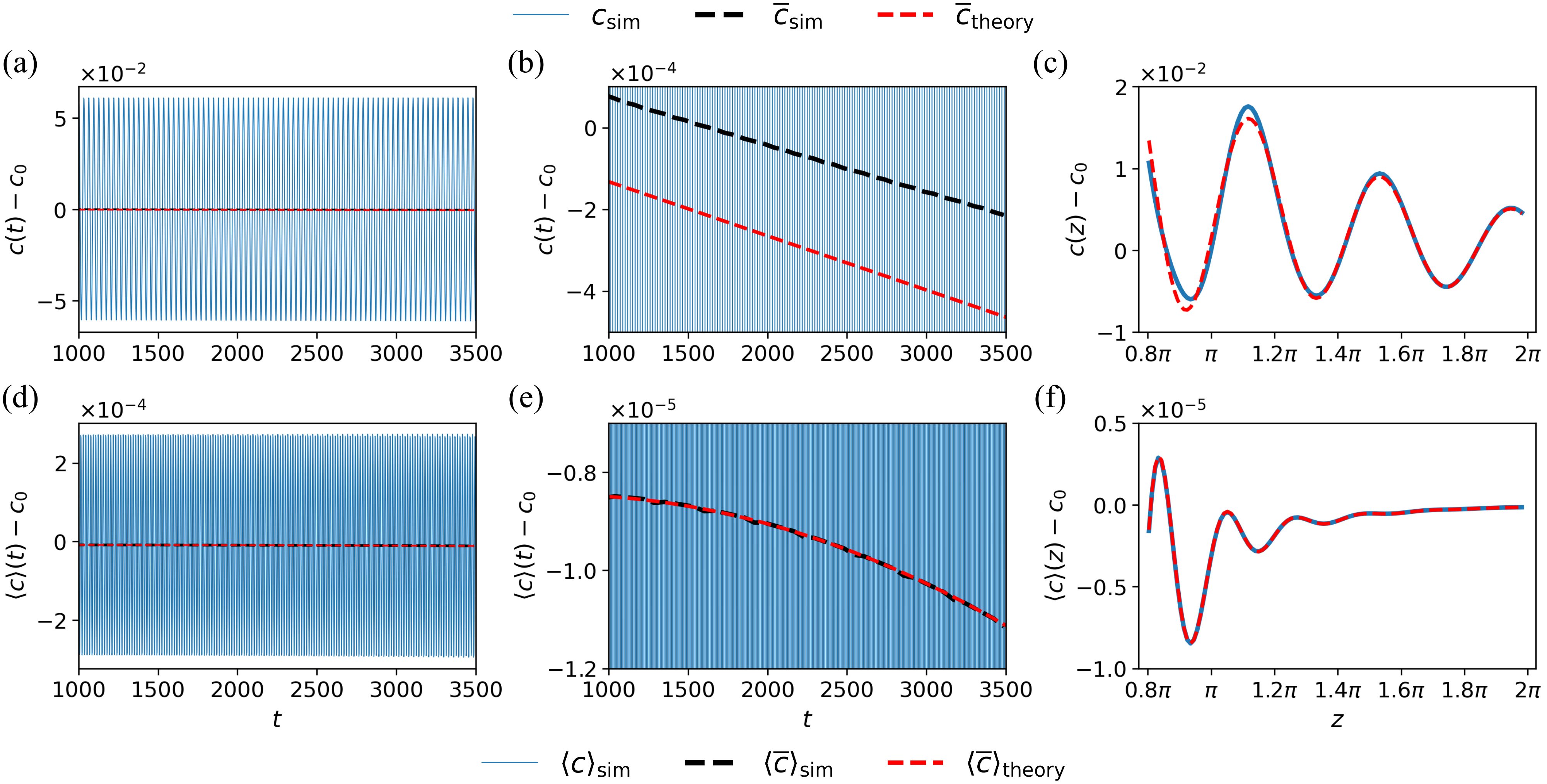}
    \caption{Transport by four waves  with the same wave frequencies $\omega_1=\omega_2=\omega_3=\omega_4$ at (a,b) $x_0=\pi$, $z_0=0.998\pi$ and (c) $x_0=\pi$, $t=4500$.
    The wave parameters are in \eqref{eq:4waves_param_same_om}.
    The second row shows net vertical transport $\langle c \rangle(z,t)-c_0$ at $z_0=0.998\pi$ (d,e) and $t=4500$ (f). The middle panels (b,e) are zoomed-in views of the left (a,d). The oscillations (blue) and the time-averaged mean transport (black) are compared with theory (red). A comparison between the fitted parameters and theoretical values is provided in Table \ref{tab}.  } 
    \label{fig:c_4waves_a}
\end{figure}

\begin{figure}[tb!]
    \centering
    \includegraphics[width=\linewidth]{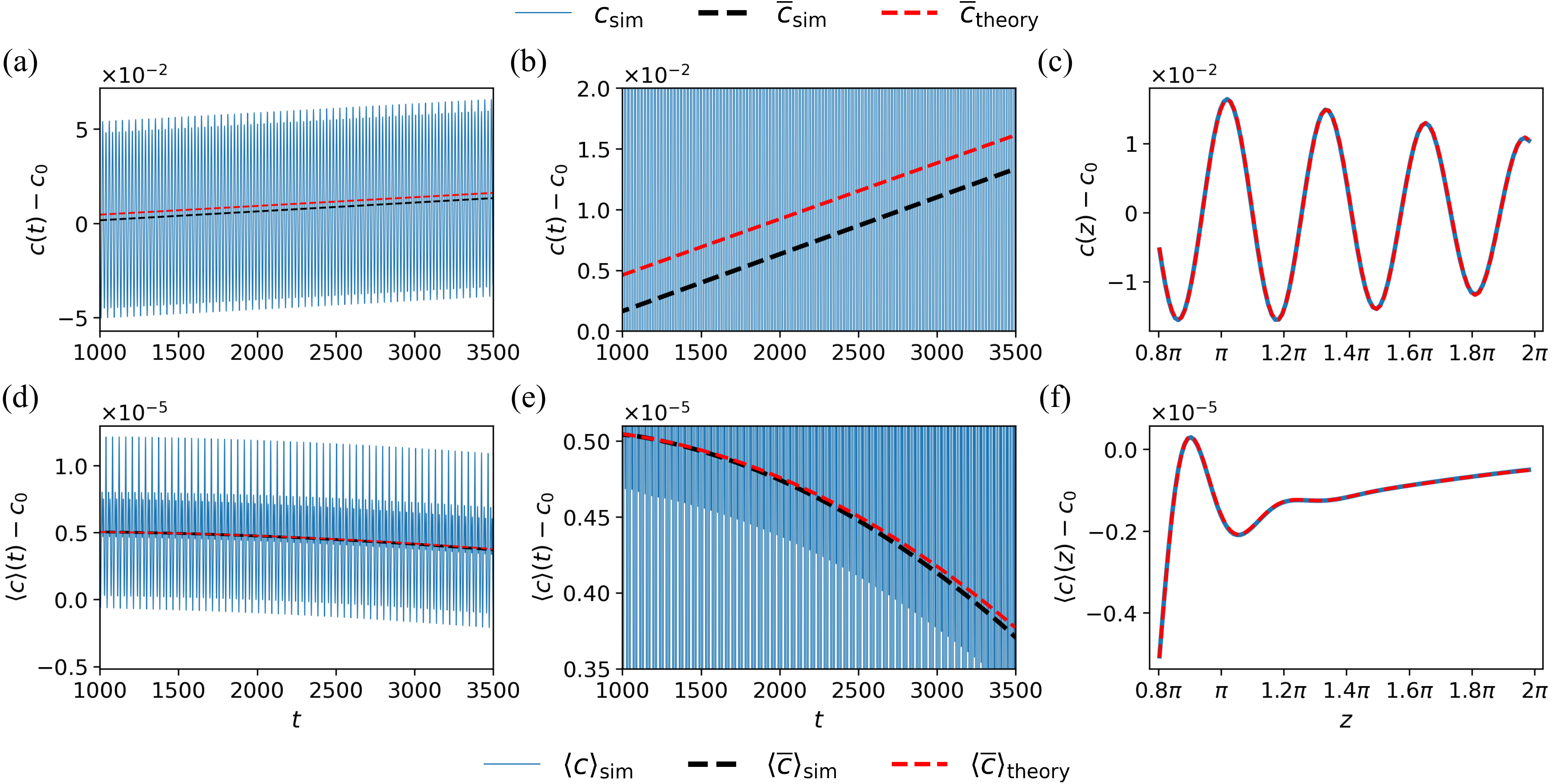}
    \caption{Transport and net vertical transport by four waves with $\omega_1=\omega_2=0.15$ and $\omega_3=\omega_4=0.3$ (blue), compared with the theoretical predictions (red). The wave parameters are given in \eqref{eq:4waves_param_same_diff1}.
    Panels (a,b,d,e) show the transport at the fixed location $x_0=\pi$, $z_0=0.998\pi$, while panels (c,f) show the profiles at $x_0=\pi$ and $t=4500$. Panels (b,e) are zoomed-in views of (a,d), respectively. The black curves show the time-averaged transport. The fitted parameters and theoretical values are reported in Table~\ref{tab}.  }
    \label{fig:c_4waves_c1}
\end{figure}

\begin{figure}[tb!]
    \centering
    \includegraphics[width=\linewidth]{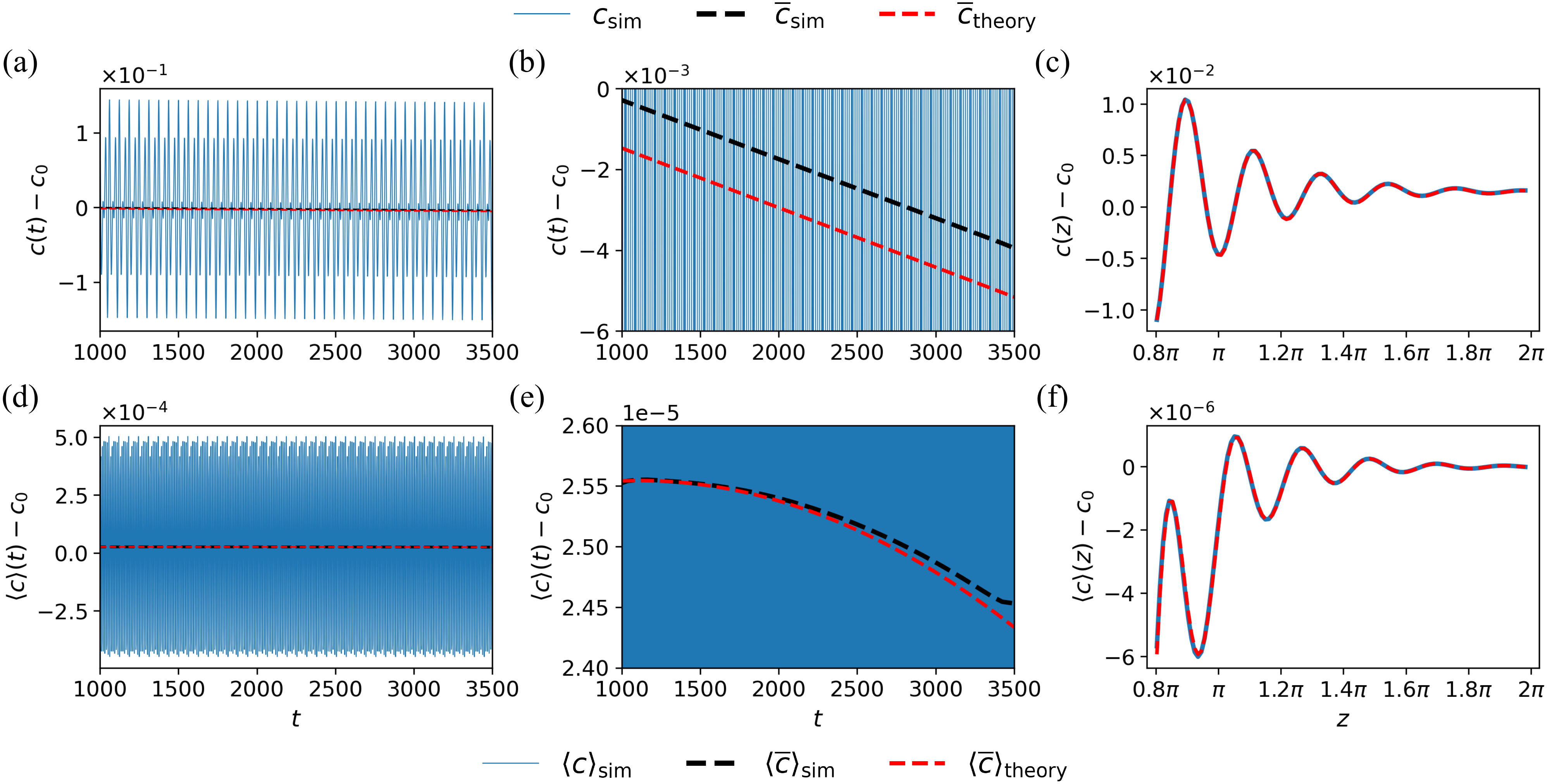}
    \caption{Transport and net vertical transport by four waves with  $\omega_1=\omega_2=0.3$ and $\omega_3=\omega_4=0.2$ compared with the theoretical predictions (red). The wave parameters are given in \eqref{eq:4waves_param_same_diff2}. Panels (a,b,d,e) show the transport at the fixed location $x_0=\pi$, $z_0=0.998\pi$, while panels (c,f) show the profiles at $x_0=\pi$ and $t=4500$. The middle panels (b,e) are zoomed-in views of the left (a,d). The black curves show the time-averaged transport. The fitted parameters and theoretical values are provided in Table \ref{tab}.  }
    \label{fig:c_4waves_c2}
\end{figure}

\begin{enumerate}[label=(\alph*)]
    \item $\omega_1=\omega_2=\omega_3=\omega_4$. When all four waves have the same frequency, resonance occurs and enables net transport over time. In Fig.~\ref{fig:c_4waves_a}, we use the following example parameters for the wave forcing at the boundary so that $\omega_j =0.2$, $j=1,2,3,4$, and 
    \begin{equation} \label{eq:4waves_param_same_om}
        \begin{aligned}
            (a_1,k_{h,1})=(0.002,2), &\quad (a_2,k_{h,2})=(0.002,-2), \\ (a_3,k_{h,3})=(0.002,-3), &\quad (a_4,k_{h,4})=(0.0005,1).
        \end{aligned}
    \end{equation}
    As in the previous cases, the net transport in both $c(t)-c_0$ and $\langle c(t) \rangle-c_0$, as well as the residual passive tracer changes $c(z)-c_0$ and $\langle c(z) \rangle-c_0$ at the final time, are in good agreement with the theoretical prediction for 4--wave interactions.

    \item $\omega_1=\omega_2=\omega_3\neq\omega_4$. When three of the four waves share the same frequency and the fourth has a different frequency, the theory predicts that the net transport is the same as for three waves, as discussed in Section \ref{sec:sim_3wave}(a). To see this, note first that the leading-order net transport $c_{1s}$ in \eqref{eq:c1s_int} is a linear superposition of the contributions from pairs of waves with the same frequency. The fourth wave, having a different frequency, does not contribute to $c_{1s}$, and therefore, does not contribute the net vertical transport $\langle c \rangle(z,t)$ as well.

    \item $\omega_1=\omega_2$, $\omega_3=\omega_4$, with $\omega_1\neq\omega_3$.
    We test two cases. In the first, $\omega_3=2\omega_1$. The results are shown in Fig.~\ref{fig:c_4waves_c1} for the following boundary-forcing wave parameters
    \begin{equation} \label{eq:4waves_param_same_diff1}
        \begin{aligned}
            (a_1,k_{h,1},\omega_1)=(0.002,2,0.15), &\quad
            (a_2,k_{h,2},\omega_2)=(0.002,-2,0.15), \\
            (a_3,k_{h,3},\omega_3)=(0.002,-3,0.3), &\quad
            (a_4,k_{h,4},\omega_4)=(0.001,1,0.3).
        \end{aligned}
    \end{equation}
    The simulation results show good agreement with the theoretical predictions for both the transport $c_{1s}$ in \eqref{eq:c1s_int} and the net vertical transport $c_{2s}$ in \eqref{eq:c2s_n}. The contribution $c_{1s}$ is the linear superposition of the transport induced by the two equal-frequency wave pairs, whereas $c_{2s}$ arises from fourth-order interactions involving all four waves under the quartet resonance condition.
    
    As shown in Table \ref{tab}, the fitted second- and fourth-order transport coefficients $\beta$ and $\gamma_1$ are also in good agreement with their theoretical values. These coefficients are obtained by fitting the averaged time evolution in Figs.~\ref{fig:c_4waves_c1}(a,b,d,e), while the fitted starting time $t_{\rm fit}$ remains close to its expected value. The final tracer profiles in Figs.~\ref{fig:c_4waves_c1}(c,f) are fitted using an effective transport time $t_{\rm transport}$, which is also in good agreement with the estimated transport duration.
    
    Figure~\ref{fig:c_4waves_c2} shows the second case, again consisting of two frequency-matched wave pairs, with parameters
    \begin{equation} \label{eq:4waves_param_same_diff2}
        \begin{aligned}
            (a_1,k_{h,1},\omega_1)=(0.002,2,0.3), &\quad
            (a_2,k_{h,2},\omega_2)=(0.001,-2,0.3), \\
            (a_3,k_{h,3},\omega_3)=(0.002,-3,0.2), &\quad
            (a_4,k_{h,4},\omega_4)=(0.002,1,0.2).
        \end{aligned}
    \end{equation}
    The simulation results again show good agreement with the theoretical predictions for both $c_{1s}$ in \eqref{eq:c1s_int} and $c_{2s}$ in \eqref{eq:c2s_n}. As shown in Table~\ref{tab}, the fitted values of $\beta$ and $\gamma_1$ agree well with theory. However, the fitted starting time $t_{\rm fit}$ is approximately twice the expected value. We attribute this discrepancy to the wave turn-on transient when the two frequencies are not integer multiples of one another. Nevertheless, the total net transport after turning off the wave forcing Fig.~\ref{fig:c_4waves_c2}(f) agrees well with theory. This suggests any effects of the turn-on and turn-off transients cancel.

     \item $\omega_1=\omega_2\neq\omega_3\neq\omega_4$. In this configuration, the net transport reduces to that of two waves with frequency $\omega_1=\omega_2$, while the additional distinct-frequency waves do not contribute.

    \item $\omega_1\neq\omega_2\neq\omega_3\neq\omega_4$. With all four frequencies distinct, the transport given by $c_{1s}$ in \eqref{eq:c1s_int} is zero, and therefore the net vertical transport is also zero.

\end{enumerate}

\clearpage
\section{Transport coefficient}
\label{sec:coeffient}

Before examining the transport under stellar conditions, we distinguish the parameter regime used in the simulations from that considered below. In the simulations, we set
$\kappa=5\cdot 10^{-4}$, nondimensionalized using $N$ and the radius of the star's convective core, whereas a representative value for a zero-age main-sequence
$15M_\odot$ star is $\kappa \sim 10^{-9}$ in the same nondimensionalization \citep[e.g.][]{Anders2023b}. 
We also note that the fitted transport coefficients $\beta$ and $\gamma_1$ in Table \ref{tab} are predominantly negative for the wave parameters used in the simulations. However, different wave combinations can produce either positive or negative transport.

We now assess which wave combinations most efficiently transport chemicals in stars.
The wave-driven transport is described by \eqref{eq:c2s_n} under the resonance conditions for the wave frequencies and wavenumbers. For two waves, $(i,j,k,l)$ take values from the set $\{1,2\}$. For three waves, both two-wave and triad interactions arise, corresponding respectively to terms containing two or three \textit{distinct} indices selected from $\{1,2,3\}$. Similarly, for four or more waves, the transport contains two-wave, triad, and quartet interactions, the last involving four \textit{distinct} indices selected from $\{1,2,3,4,\dots,n\}$. 

For example, consider two waves $(A_n,k_{h,n},\omega)$, $n=1,2$, with identical frequencies.
To satisfy the horizontal wavenumber resonance condition in \eqref{eq:c2s_phase}, the admissible index combinations are
\begin{equation*}
    (i,j,k,l) \in \mathcal{S}, \quad \text{where} ~ \mathcal{S} = \{ (1,2,2,1), ~ (2,1,1,2) \}.
\end{equation*}
For triad interactions, we require identical wave frequencies together with the horizontal resonance condition in \eqref{eq:c2s_phase}. Without loss of generality, we can take $2k_{h,2}=k_{h,1}+k_{h,3}$, and the admissible index combinations are
\begin{equation*}
    (i,j,k,l) \in \mathcal{S}, \quad \text{where} ~ \mathcal{S} = \{ (1,2,3,2), ~ (3,2,1,2), ~ (2,1,2,3),  ~ (2,3,2,1) \}.
\end{equation*}
Quartet interactions with distinct $(i,j,k,l)$ occur among four wave sets $(A_n,k_{h,n},\omega_n)$, $n=1,2,3,4$.
Assume the four waves satisfy $k_{h,1}-k_{h,2}+k_{h,3}-k_{h,4}=0$.
The set of resonant permutations depends on the frequency pairing. 
When all four frequencies are equal $\omega_1=\omega_2=\omega_3=\omega_4$, we have 
\begin{equation*}
    \begin{aligned}
        (i,j,k,l) \in \mathcal{S}, \quad \text{where} ~ \mathcal{S} = \{ &(1,2,3,4), ~ (3,2,1,4), ~ (1,4,3,2), ~ (3,4,1,2), \\
        & (2,1,4,3), ~ (4,1,2,3), ~ (2,3,4,1), ~ (4,3,2,1) \}.
    \end{aligned}
\end{equation*} 
For the case $\omega_1=\omega_2\neq\omega_3=\omega_4$, the admissible permutations are
\begin{equation}
        (i,j,k,l) \in \mathcal{S}, \quad \text{where} ~ \mathcal{S} = \{ (1,2,3,4), ~ (3,4,1,2), ~ (2,1,4,3),  ~ (4,3,2,1) \}.
\end{equation}
For a system of $n>4$ waves, the quartet contributions are obtained by considering the $\binom{n}{4}$ combinations of four distinct waves and retaining those that satisfy the resonance conditions.

In the following, we examine the net vertical transport $\langle \overline{c} \rangle$ induced by two-wave, triad, and quartet interactions, separately, in order to determine which interaction class contributes most strongly to the transport. Within each interaction class, multiple resonant permutations of the indices $(i,j,k,l)$ may correspond to the same phase relation, reducing the phase in \eqref{eq:c2s_phase} to a single reference phase $\overline{\Delta}$, up to a sign, i.e. $\overline{\Delta}=\pm \Delta$. However, the sine term in \eqref{eq:c2s_n} should take into account the reversal of the sign $\sin(-\Delta)=-\sin(\Delta)$.
We therefore define the permutation-summed interaction coefficients as
\begin{align}
    \overline{\alpha}_1 =  \sum_{(i,j,k,l)\in\mathcal{S}} \alpha_1, 
    \quad
    \overline{\alpha}_2  =  \sum_{(i,j,k,l)\in\mathcal{S}} \delta_{ijkl} \,\alpha_2,   \quad 
    \delta_{ijkl} = \begin{cases}
        +1 , & \overline{\Delta} = \Delta \\
        -1 , & \overline{\Delta} = -\Delta
    \end{cases}.
\end{align}
For each interaction class, the permutation-summed long-time net vertical transport \eqref{eq:c2s_n} can then be rewritten as 
\begin{subequations}  
    \begin{align} \label{eq:coeff_dc}
        \Delta  \langle \overline{c} \rangle (z,\bar{t})  &= 
        \frac{t^{2}}{2} \, a_i a_j a_k a_l \, \overline{\alpha} \,  \partial_z  \left[ \frac{d c_0}{dz} \, e^{-z\left(\frac{1}{\mu_i}+\frac{1}{\mu_j}+\frac{1}{\mu_k}+\frac{1}{\mu_l}\right)} 
        \left\langle \cos ( \overline{\Delta} - \psi_0 )
        \right\rangle  \right],
    \end{align}
where $\overline{\alpha}$ is defined non-negative as
    \begin{align}
        \overline{\alpha}=
        \begin{cases}
            |\overline{\alpha}_1|, \\
            (\overline{\alpha}_1^2 + \overline{\alpha}_2^2)^{1/2},
        \end{cases}
        \quad \text{and} \quad 
        \psi_0=
        \begin{cases}
            \arg(\overline{\alpha}_1), & \text{2--wave interaction with $\overline{\Delta}=\Delta=0$} , \\
            \arg(\overline{\alpha}_1+i\overline{\alpha}_2), & \text{otherwise}.
        \end{cases}
    \end{align}
\end{subequations}
Note that the amplitude product $a_i a_j a_k a_l$ and the decay rate
$1/\mu_i+1/\mu_j+1/\mu_k+1/\mu_l$ are invariant over
$\mathcal{S}$ in each class. Therefore, in \eqref{eq:coeff_dc}, $(i,j,k,l)$ may be chosen as any element of $\mathcal{S}$.

First, we nondimensionalize the wave variables using the length and time scales $L=|k_{h,1}|^{-1}$ and $T=N^{-1}$ such that 
$$
    \hat{t} = N t, \quad \hat{z}=|k_{h,1}|z,  \quad \hat{\mathbf{k}}_p= \mathbf{k}_p/|k_{h,1}|, \quad \hat{\omega}_p = \omega_p/N, \quad \hat{\kappa} = \kappa \, k_{h,1}^2/N, \quad \hat{\mu}_p=|k_{h,1}|\mu_p,
$$
and 
$$
    \hat a_p= a_p \, |k_{h,1}|/N, \quad \hat{\overline{\alpha}} = \overline{\alpha} \, N^2/k_{h,1}^2.
$$
Hats denote nondimensional quantities and are omitted hereafter for simplicity.
From \eqref{eq:coeff_dc}, we now define the nondimensional effective transport coefficient as 
\begin{equation}
    D_c =  a_i a_j a_k a_l \, \overline{\alpha}  . \label{eq:coeff_a_alpha}
\end{equation}
For waves to propagate for more than one period, they must not be strongly nonlinear.
That is, 
\begin{equation} \label{eq:transport_ap}
    a_p \lesssim \omega_p/|k_{z,p}|, \quad p\in\{i,j,k,l\}.
\end{equation}
Because the amplitudes for which waves propagate varies with $\omega$ and $k_h$, we will estimate $D_c$ for the highest possible amplitude waves, i.e., setting $a_p=\omega_p/|k_{z,p}|$.
This gives an upper limit for $D_c$.

Our objective is to determine how the transport depends on the horizontal wavenumbers, wave frequency, and radiative diffusivity. We therefore seek a scaling relation of the form
\begin{equation} \label{eq:coeff_scale}
    D_c \simeq d_c \, F \left( k_{h,i}, k_{h,j}, k_{h,k}, k_{h,l} \right) \omega^m \kappa^n,
\end{equation}
where the constant $d_c$, the wavenumber--dependent function $F$, and the exponents $m$ and $n$ are determined below. 
While $D_c$ has power-law dependencies on $\omega$ and $\kappa$, it depends more weakly on $k_h$. As shown below, it increases somewhat with $k_h$. However, we focus on the case of $k_h\sim \mathcal{O}(1)$ as waves with larger $k_h$ need higher frequencies, which correspond to weaker transport ($m<0$), and are also less efficiently excited by convection. Table~\ref{tab2} summarizes the dependence on $\omega$ and $\kappa$ for different types of wave interactions.

\begin{table}
\hspace{-0.2in}
\begin{tabular}{|c|c|c|c|c|c|c|}
\hline
Interaction class & Fig. & $k_{h,2}/k_{h,1}$ & $k_{h,3}/k_{h,1}$ & $m$ & $n$ & $d_c$ \\
\hline

\multirow{2}{*}{Two-wave} & \multirow{2}{*}{8(b,c)} 
& $-2$ & N/A & $-2$ & $2$ & $4.50$ \\
\cline{3-7}
&
& $2$ & N/A & $-8$ & $4$ & $7.03$ \\
\hline

\multirow{2}{*}{Triad}
& \multirow{2}{*}{9(b,c)}
& $-2$ & $-5$ & $-5$ & $3$ & $152.26$ \\
\cline{3-7}
&
& $2$ & $3$ & $-8$ & $4$ & $60.96$ \\
\hline

\multirow{2}{*}{Quartet with $\omega_1=\omega_2=\omega_3=\omega_4$}
& \multirow{2}{*}{10(c,d)}
& $-2$ & $-1/2$ & $-2$ & $2$ & $17.80$ \\
\cline{3-7}
&
& $2$ & $-1/2$ & $-2$ & $2$ & $6.75$ \\
\hline

\multirow{2}{*}{Quartet with $\omega_1=\omega_2\neq\omega_3=\omega_4$}
& \multirow{2}{*}{11(c,d,e)}
& $-3$ & $-1$ & $-2$ & $2$ & $35.97$ \\
\cline{3-7}
&
& $6$ & $2$ & $-5$ & $3$ & $194.13$ \\
\hline

\end{tabular}
\caption{Frequency and diffusivity scalings of the transport coefficient
$D_c \sim d_c \,\omega^m \kappa^n$ at selected wavenumber ratios $k_{h,2}/k_{h,1}$ and $k_{h,3}/k_{h,1}$ for different interaction classes.}
\label{tab2}
\end{table}

\begin{figure}[tb!]
    \centering
    \includegraphics[width=\linewidth]{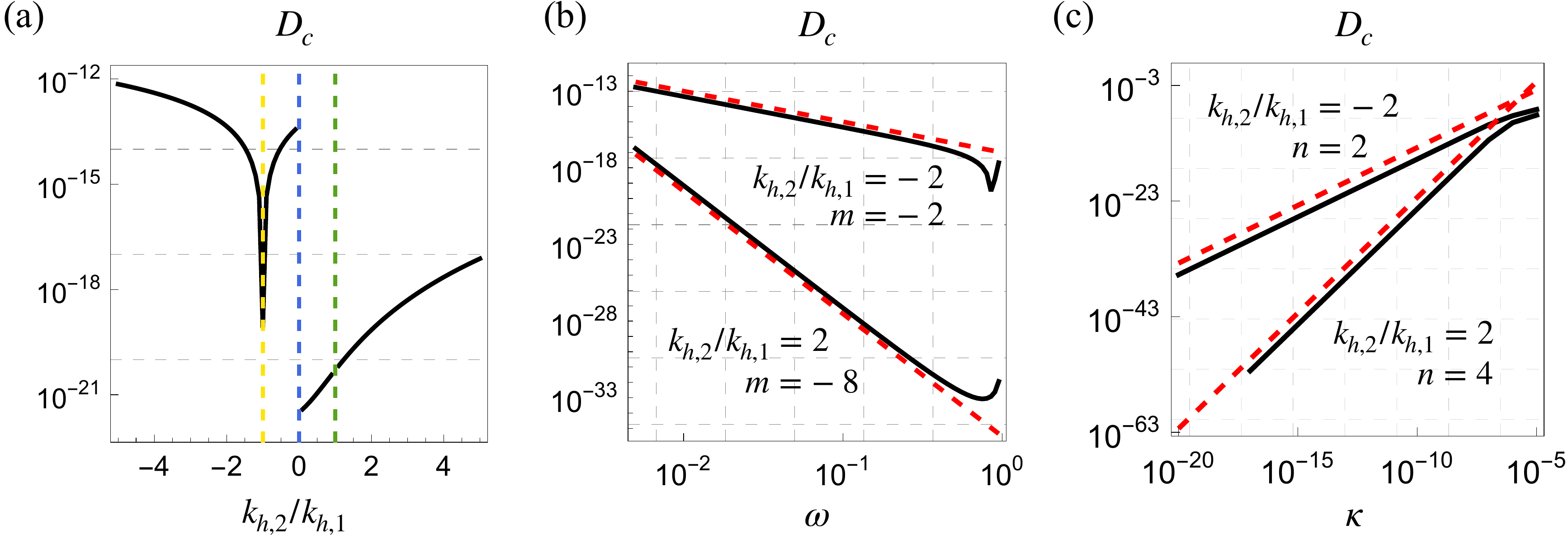}
    \caption{Nondimensional transport coefficient $D_c = \prod_{p\in\{i,j,k,l\}}a_p\cdot\bar{\alpha}  $ from \eqref{eq:coeff_a_alpha} for two-wave interactions. (a) Variation with wavenumber $k_{h,2}/k_{h,1}$ for $\omega=10^{-2}$ and $\kappa=10^{-9}$. The blue, green, and yellow dashed lines indicate $k_{h,2}=0$, $k_{h,1}=k_{h,2}$, and $k_{h,1}=-k_{h,2}$, respectively. (b) Frequency scaling and (c) diffusivity scaling for fixed $k_{h,2}/k_{h,1}=\pm2$.}
    \label{fig:alpha_duo}
\end{figure}

\begin{figure}[tb!]
    \centering
    \includegraphics[width=\linewidth]{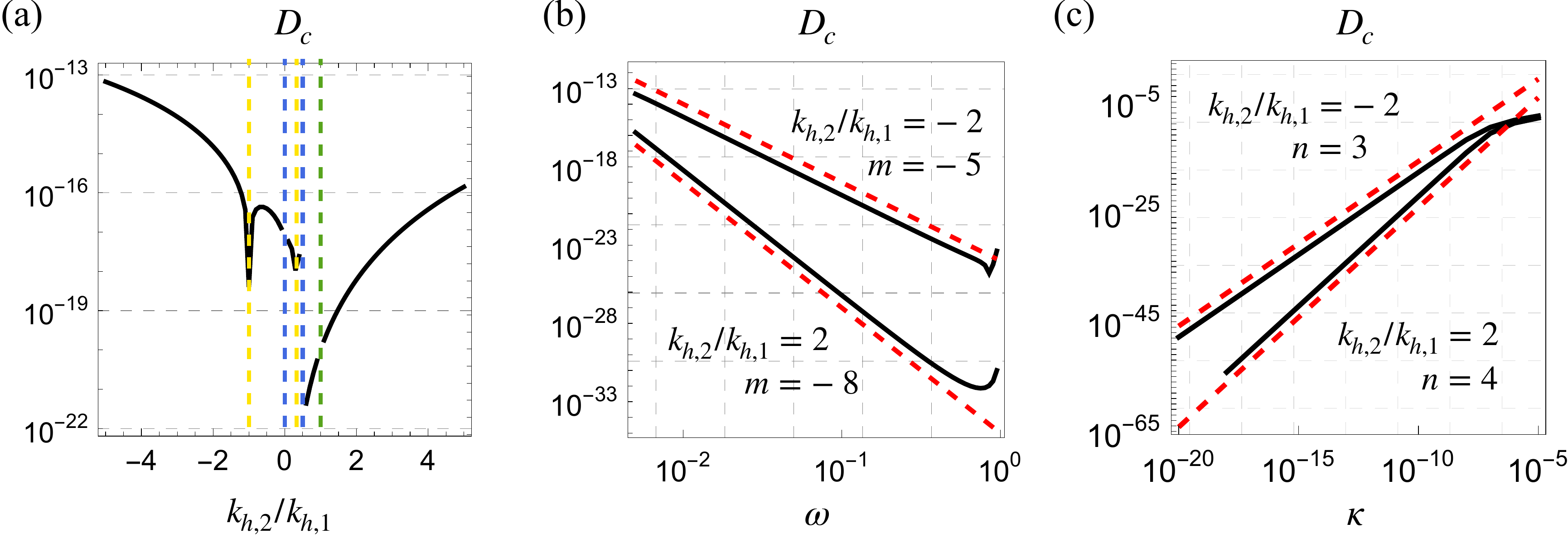}
    \caption{Nondimesional transport coefficient \eqref{eq:coeff_a_alpha} for triad interactions. (a) Variation with wavenumber $k_{h,2}/k_{h,1}$ for $\omega=10^{-2}$ and $\kappa=10^{-9}$. The blue dashed lines indicate $k_{h,2}=0$ and $k_{h,3}=0$, the green dashed line indicates $k_{h,1}=k_{h,2}=k_{h,3}$, and the yellow dashed lines indicate $k_{h,1}=-k_{h,2}$ and $k_{h,1}=-k_{h,3}$. (b,c) Frequency and diffusivity scalings for fixed $k_{h,2}/k_{h,1}=\pm2$. }
    \label{fig:alpha_triad}
\end{figure}

Figure \ref{fig:alpha_duo} shows the transport coefficient for the two-wave interaction between two weakly nonlinear waves $(a_1,k_{h,1},\omega_1)$ and $(a_2,k_{h,2},\omega_2)$. 
Figure \ref{fig:alpha_duo}(a) illustrates the dependence of \eqref{eq:coeff_a_alpha} on wavenumber $k_{h,2}/k_{h,1}$ for fixed $\omega=\omega_1=\omega_2=10^{-2}$ and $\kappa=10^{-9}$.
A jump discontinuity occurs at $k_{h,2}=0$, while a removable singularity occurs at $k_{h,2}=k_{h,1}$, where the system degenerates into a single-wave state and no transport is expected. The transport coefficient attains a local minimum at $k_{h,2}/k_{h,1}=-1$, corresponding to a pair of counter-propagating waves.
Overall, the transport coefficient tends to increase when $k_{h,2}$ is opposite in sign to $k_{h,1}$ and has a larger magnitude. 
Next, we examine the frequency dependence of the transport coefficient \eqref{eq:coeff_a_alpha} for $k_{h,2}/k_{h,1}=-2$ and $k_{h,2}/k_{h,1}=2$. With $\kappa=10^{-9}$ held fixed, Fig.~\ref{fig:alpha_duo}(b) shows that the coefficient scales as $D_c\sim\omega^{m}$, with $m=-2$ and $m=-8$ for the former and latter cases, respectively. We then set $\omega=10^{-2}$ and investigate the dependence on the radial diffusivity $\kappa$. Figure \ref{fig:alpha_duo}(c) shows that the coefficient scales as $D_c \sim \kappa^{n}$, with $n=2$ and $n=4$, respectively. The corresponding prefactors are $d_c=4.50$ and $d_c=7.03$, respectively.

We next consider the triad interaction among three distinct waves, $(a_1,k_{h,1},\omega_1)$, $(a_2,k_{h,2},\omega_2)$, and $(a_3,k_{h,3},\omega_3)$. As discussed above, these waves must have the same frequency, $\omega=\omega_i$, $i=1,2,3$, and satisfy the resonance condition given by \eqref{eq:c2s_phase}. Without loss of generality, we will assume $2k_{h,2}=k_{h,1}+k_{h,3}$. 
For fixed $\omega=10^{-2}$ and $\kappa=10^{-9}$, Fig.~\ref{fig:alpha_triad}(a) shows the dependence of the transport coefficient on the wavenumbers. 
Jump discontinuities occur at $k_{h,2}=0$ and $k_{h,3}=0$, while a removable singularity occurs at $k_{h,1}=k_{h,2}=k_{h,3}$, where the three-wave system degenerates into a single wave and no transport is expected. Local minima occur at $k_{h,2}=-k_{h,1}$ and $k_{h,2}=-k_{h,3}$, corresponding to the situation in which two of the waves are counter-propagating with equal frequencies and wavenumber magnitudes.
We further examine two cases $k_{h,2}/k_{h,1}=-2$ and $k_{h,2}/k_{h,1}=2$. With $\kappa=10^{-9}$ fixed, Fig.~\ref{fig:alpha_triad}(b) shows that the transport coefficient scales as $D_c\sim\omega^m$, with $m=-5$ and $m=-8$, respectively. With $\omega=10^{-2}$ fixed, Fig.~\ref{fig:alpha_triad}(c) shows that the coefficient scales as $D_c\sim\kappa^n$, with $n=3$ and $n=4$, respectively.
The prefactors are $d_c=152.26$ and $d_c=60.96$.
We find that, for both two-wave and triad interactions, larger transport coefficients are associated with counter-propagating wave pairs of larger horizontal wavenumber.

\begin{figure}[tb!]
    \centering
    \includegraphics[width=\linewidth]{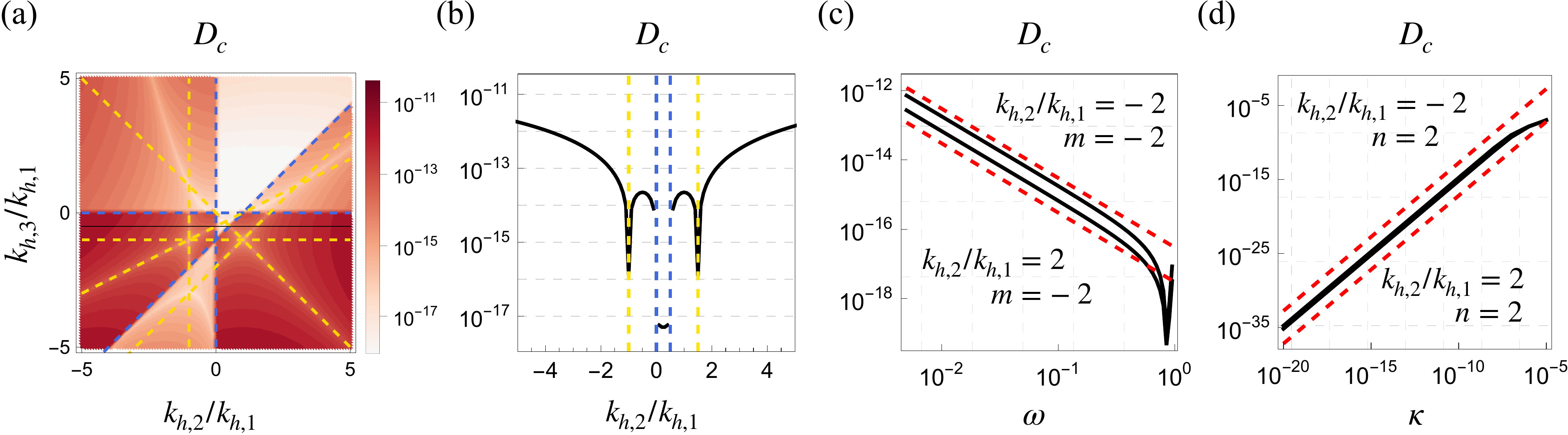}
    \caption{Nondimensional transport coefficient \eqref{eq:coeff_a_alpha} for quartet interactions among four distinct waves of common frequency $\omega$. (a) Variation with wavenumber ratios $k_{h,2}/k_{h,1}$ and $k_{h,3}/k_{h,1}$ at $\omega=10^{-2}$ and $\kappa=10^{-9}$. (b) One-dimensional slice at $k_{h,3}/k_{h,1}=-1/2$ (the black line in (a)). The blue lines indicate $k_{h,2}=0$, $k_{h,3}=0$, or $k_{h,4}=0$, and the yellow lines indicate wave sets with opposite horizontal wavenumbers. (c) Frequency and diffusivity scalings for selected wavenumbers $k_{h,2}/k_{h,1}=-2$ and $k_{h,2}/k_{h,1}=2$, yielding $m=-2$ and $n=2$ in both cases.}
    \label{fig:alpha_quartet_same}
\end{figure}

\begin{figure}[tb!]
    \centering
    \includegraphics[width=\linewidth]{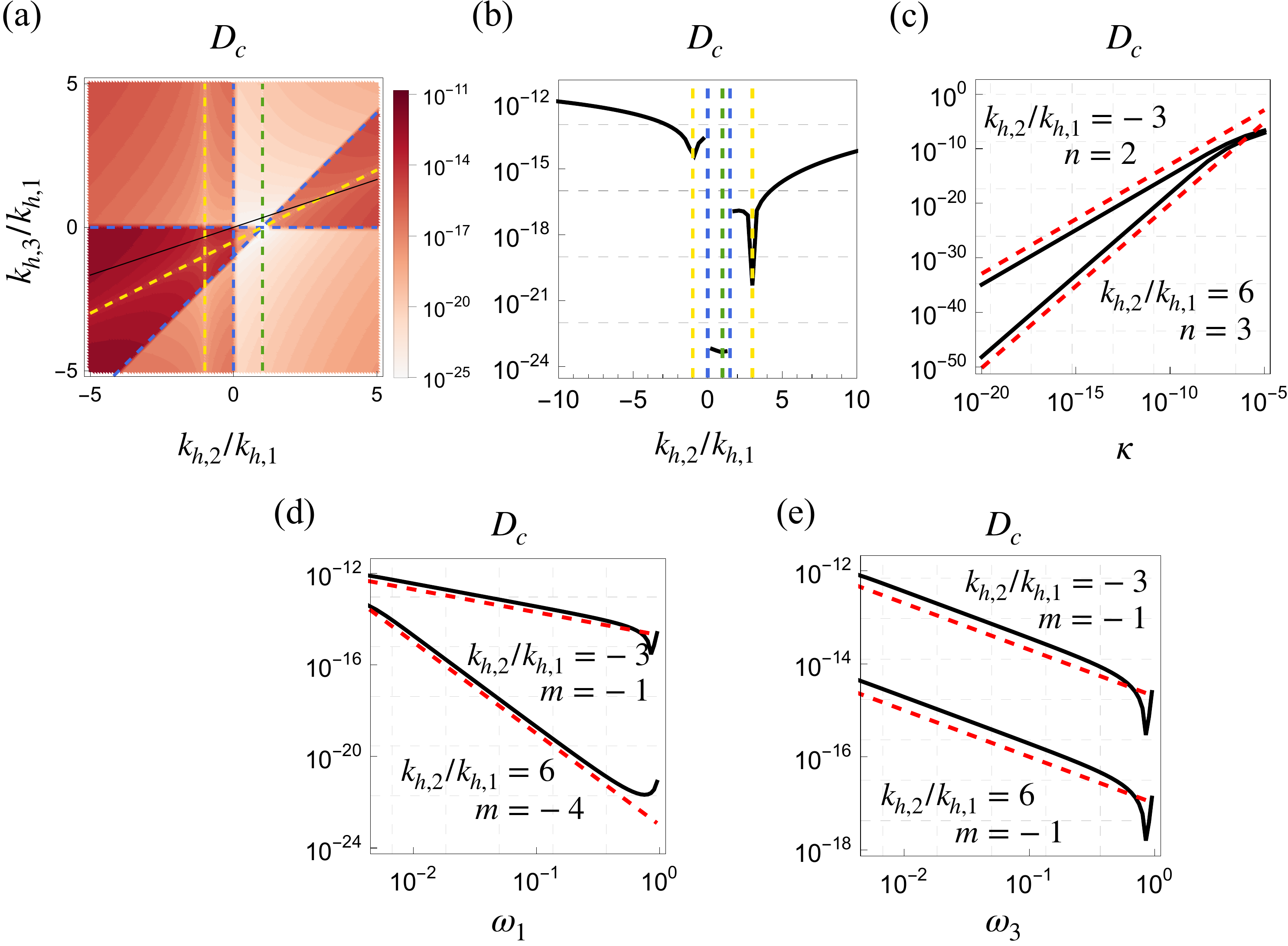}
    \caption{Nondimensional transport coefficient \eqref{eq:coeff_a_alpha} for quartet interactions among four distinct waves with two frequencies $(\omega_1,\omega_3)$. (a) Variation with wavenumber ratios $k_{h,2}/k_{h,1}$ and $k_{h,3}/k_{h,1}$ for $\omega_1=10^{-2}$, $\omega_3=3\times 10^{-2}$, and $\kappa=10^{-9}$.
    The blue lines indicate $k_{h,2}=0$, $k_{h,3}=0$, or $k_{h,4}=0$; the green line indicates the degenerate wave sets with $k_{h,1}=k_{h,2}$ and $k_{h,3}=k_{h,4}$; and the yellow lines indicate wave sets with opposite horizontal wavenumbers.
    (b) One-dimensional slice at $k_{h,3}/k_{h,2}=1/3$ (the black line in (a)). (c) Radial diffusivity scaling for $k_{h,2}/k_{h,1}=-3$, resulting in $n=2$, and for  $k_{h,2}/k_{h,1}=6$, resulting in $n=3$. (d) Frequency scaling with respect to $\omega_1$, giving $m=-1$ and $m=-4$ for $k_{h,2}/k_{h,1}=-3$ and $6$, respectively. (e) Frequency scaling with respect to $\omega_3$, giving $m=-1$.}
    \label{fig:alpha_quartet_diff}
\end{figure}

In quartet interactions involving four distinct waves, by non-dimensionalizing with $k_{h,1}=1$, only two independent parameters $k_{h,2}/k_{h,1}$ and $k_{h,3}/k_{h,1}$ remain, while $k_{h,4}$ is determined by the resonance condition $k_{h,4}=k_{h,1}-k_{h,2}+k_{h,3}$. 
Figure \ref{fig:alpha_quartet_same} shows the transport coefficient \eqref{eq:coeff_a_alpha} for the case in which all four waves have the same frequency $\omega=\omega_i$, $i=1,2,3,4$. For fixed $\omega=10^{-2}$ and $\kappa=10^{-9}$, Fig.~\ref{fig:alpha_quartet_same}(a) shows the dependence of the transport coefficient on $k_{h,2}/k_{h,1}$ and $k_{h,3}/k_{h,1}$. The largest transport magnitudes occur when $k_{h,3}/k_{h,1}<0$. Motivated by this trend, we examine a one-dimensional slice at $k_{h,3}/k_{h,1}=-1/2$ in Fig.~\ref{fig:alpha_quartet_same}(b). 
The blue lines in Fig.~\ref{fig:alpha_quartet_same}(a) are the degenerate cases $k_{h,2}=0$, $k_{h,3}=0$, or $k_{h,4}=0$, for which there is no internal gravity wave, and hence the quartet transport coefficient is zero. As a horizontal wavenumber changes sign across these lines, the wave polarization also changes sign, producing the jump discontinuities in Fig.~\ref{fig:alpha_quartet_same}(b).
The yellow lines correspond to local minima of the transport coefficient and occur when a pair of waves is counter-propagating: (i) $k_{h,1}=-k_{h,2}$; (ii) $k_{h,1}=-k_{h,3}$ and $k_{h,2}=-k_{h,4}$; (iii) $k_{h,1}=-k_{h,4}$; (iv) $k_{h,2}=-k_{h,3}$; and (v) $k_{h,3}=-k_{h,4}$.
The transport coefficient increases with the magnitudes of $k_{h,2}$ and $k_{h,3}$.
For fixed $k_{h,2}/k_{h,1}=\pm2$ and $k_{h,3}/k_{h,1}=-1/2$, Figs.~\ref{fig:alpha_quartet_same}(c) and (d) show that the transport coefficient scales as $D_c\sim\omega^{-2}\kappa^{2}$, while the prefactors are $d_c=17.80$ for $k_{h,2}/k_{h,1}=-2$ and $d_c=6.75$ for $k_{h,2}/k_{h,1}=2$.

We turn to quartet interactions involving two distinct frequencies, with $\omega_1=\omega_2$ and $\omega_3=\omega_4$. At $\omega_3=3\omega_1=3\cdot10^{-3}$, Fig.~\ref{fig:alpha_quartet_diff}(a) shows the transport coefficient in the $(k_{h,2}/k_{h,1},k_{h,3}/k_{h,1})$ plane. The largest transport magnitudes occur in two regions near the diagonal. To characterize these regions, we fix $k_{h,3}/k_{h,1}=1/3$ and plot the transport coefficient against $k_{h,2}/k_{h,1}$ in Fig.~\ref{fig:alpha_quartet_diff}(b). 
Jump discontinuities occur at $k_{h,2}=0$, $k_{h,3}=0$, and $k_{h,4}=0$. Removable singularities occur at $k_{h,1}=k_{h,2}$ and $k_{h,3}=k_{h,4}$, where the quartet interaction degenerates into two-wave interaction with different frequencies, and therefore produces no transport. Local minima occur at $k_{h,1}=-k_{h,2}$ and $k_{h,3}=-k_{h,4}$, corresponding to counter-propagating wave pairs.
Figure \ref{fig:alpha_quartet_diff}(c) indicates the scaling $D_c\sim\kappa^{n}$, with $n=2$ for $k_{h,2}=-3k_{h,1}$ and $n=3$ for $k_{h,2}=6k_{h,1}$. We then fix these wavenumbers with $\kappa=10^{-9}$ and vary the frequencies. Figures \ref{fig:alpha_quartet_diff}(d,e) show that $D_c\sim\omega_1^{-1}\omega_3^{-1}$ for $k_{h,2}=-3k_{h,1}$, and $D_c\sim\omega_1^{-4}\omega_3^{-1}$ for $k_{h,2}=6k_{h,1}$.
Notice that the scalings with respect to $\omega_1$ and $\omega_3$ can be different. This asymmetry is attributed to the stronger frequency dependence of wave pairs with opposite-signed wavenumbers compared with those of the same sign. 
The corresponding prefactors are $d_c=35.97$ and $d_c=194.13$, respectively.

In summary, two-wave, triad, and quartet interactions produce transport coefficients of comparable magnitude, with no interaction class appearing significantly more important than the others. A common feature across all cases is that enhanced transport is associated with at least one wave propagating in the direction opposite to the others. In the weakly damped regime $\omega\gg \kappa^{1/3}$, the dominant contribution to the net vertical transport is given by
\begin{align}  \label{eq:transport_final}
\Delta  \langle \overline{c} \rangle (z,\bar{t}) 
&\simeq 
\frac{t^2}{2} \, d_c\, \omega^{-2} \kappa^{2}  \, F \left( k_{h,i}, k_{h,j}, k_{h,k}, k_{h,l} \right)
\partial_z  \left[e^{-z\left(\frac{1}{\mu_i}+\frac{1}{\mu_j}+\frac{1}{\mu_k}+\frac{1}{\mu_l}\right)} 
\left\langle \cos ( \overline{\Delta} - \psi_0 )
\right\rangle \frac{dc_0}{dz} \right], 
\end{align}
where $d_c=\mathcal{O}(1)$ is a constant prefactor. 
For a wave field containing $n$ waves, the indices $(i,j,k,l)$ are selected from $\{1,2,3,4,\dots,n\}$ such that the corresponding waves satisfy the resonance conditions. Repeated indices are allowed.

\section{Discussion} \label{sec:discussion}

In this work, we investigated the transport of a (passive) chemical species by coherent, weakly nonlinear IGWs using both multiscale asymptotic analysis and numerical simulations. Our analysis shows that the leading-order contribution to the pointwise transport arises from the Eulerian mean velocity and Stokes drift in the presence of diffusivity. In the absence of diffusivity, the transport associated with the Eulerian mean velocity is canceled by the Stokes drift, due to the conservation of pseudo-momentum. With diffusivity, we find that the pointwise change in the tracer scales as $\overline{c}-c_0 =\mathcal{O}(a^2 t)$, for $t \ll 1/a^2$.
Net vertical transport of the tracer field arises only at fourth order in wave amplitude, scaling as $\langle \overline{c} \rangle-c_0 =\mathcal{O}(a^4 t^2)$, see (\ref{eq:c2s_n}, \ref{eq:c2s_phase}). 
The diffusivity is time-dependent and proportional to $\langle(w^S+w^E)^2\rangle$, see \eqref{eq:c2s}.
Induced two-wave transport occurs only between waves of the same frequency. Triad resonances require a common frequency and satisfy the resonance condition $2k_{h,2}=k_{h,1}+k_{h,3}$. Quartet resonances satisfy the resonance condition $k_{h,1}-k_{h,2}+k_{h,3}-k_{h,4}=0$, with either $\omega_1=\omega_2$ and $\omega_3=\omega_4$, or all four waves sharing the same frequency.

We use Dedalus to solve the nonlinear two-dimensional Boussinesq equations coupled to a tracer transport equation, with wave generation forced at the bottom boundary. By examining the long-time change in the tracer field and comparing the difference between the initial and final states, we find good agreement with our asymptotic predictions for both pointwise transport and net vertical transport.

The transport coefficient analysis compares the contributions from induced two--wave transport, triad resonances, and quartet resonances. 
Across the different interaction classes, the transport coefficients are of comparable magnitude, with no single interaction class consistently dominating. 
To place upper bounds on the transport by weakly nonlinear waves, we assumed the wave amplitude is $\omega/k_z$, which is the maximum weakly nonlinear wave amplitude. This gives an upper bound on the diffusion coefficient of $D_c\sim\omega^{-2}\kappa^2$, see \eqref{eq:transport_final}.

The present study focuses on waves generated at the bottom boundary, representing excitation by core convection and subsequent upward propagation. 
However, some waves may reflect and propagate downward.
Some stars also exhibit standing modes, which are the superposition of upward- and downward-propagating waves.
For downward-propagating waves, the corresponding weakly damped branch has $k_z>0$ and $\mu<0$. The magnitude of the transport coefficient is found to be the same as that of the corresponding upward-propagating weakly damped solution ($k_z<0$, $\mu>0$). 
In addition, 
the transport coefficients reported here represent magnitudes. However, the net tracer transport can be positive or negative depending on the wave parameters and the relative phases.

In stars, convection is expected to excite a wide spectrum of waves \citep[e.g.,][]{Goldreich1990}, with the excitation most efficient for waves matching the horizontal scale and frequency of convective motions.
For massive stars, this corresponds to low wavenumber ($\ell\sim 1$) and low frequency waves.
While these waves are predicted to be most efficiently excited, they also experience the strongest damping.

Stochastically excited waves that transport chemical species as described here are expected to have uniformly-distributed phases $\Delta$ (\ref{eq:c2s_phase}), and thus the net vertical transport by any given wave packet is equally likely to be positive or negative.
This suggests that the evolution of $\langle\overline{c}\rangle$ is expected to be diffusive over evolutionary timescales, as the net vertical transport from different wave packets cancels to leading order.

To estimate this diffusivity, we assume that each wave packet is coherent for a convective time $\tau_c$. We next take a WKB approximation where we assume the $\tilde{k} H_c\gg 1$, where $\tilde{k} = k_{z,i}-k_{z,j}+k_{z,k}-k_{z,l}$ and $H_c=c_0/(dc_0/dz)$.
Then the magnitude of the change in $c_0$ from a single wave packet of four resonant waves is
\begin{equation}
    \left|\Delta \langle \overline{c_0}\rangle\right| = \frac{\tau_c^2}{2} a^4 D_c \tilde{k} \partial_z \langle \overline{c_0}\rangle.
\end{equation}
Then averaging over many wave packets with random phases, we are left with the diffusion equation for the horizontal average of $c_0$
\begin{equation}\label{eq:diffusion}
    \partial_t \langle \overline{c_0}\rangle = \partial_z \left(D_w \partial_z\langle \overline{c_0}\rangle\right),
\end{equation}
with wave diffusivity
\begin{equation}
    D_w = \frac{\tau_c^3}{4} a^8 D_c^2 \tilde{k}^2.
\end{equation}
This has a very strong dependence on the wave amplitude, $D_w= \mathcal{O}(a^8)$. 
Above we also showed that $D_c\sim \kappa^2$, so this wave diffusivity scales like $\kappa^4$.

The maximum possible wave amplitude is $w=\omega/k_z$, because at this wave amplitude the waves are strongly nonlinear and will break within a period.
Using this wave amplitude for $a$ to get an upper bound for $D_c$, we find $D_c\sim d_c \omega^{-2}\kappa^2$, with $d_c$ an order unity prefactor. This gives an upper bound on the magnitude of the change in $c_0$ from a single wave packet of
\begin{equation}
        \left|\Delta \langle \overline{c_0}\rangle\right| \lesssim \frac{\tau_c^2}{2} d_c \omega^{-2}\kappa^2 (k_h N)^2 \tilde{k} \partial_z \langle \overline{c_0}\rangle.
\end{equation}
For a $15M_\odot$ ZAMS model, \citet{Anders2023b} calculated $\tau_c N \sim 2000$, and the highest amplitude waves that are not strongly damped have $k_h\sim 1/R_c$ and $\omega/N=10^{-3}$. We estimate $\tilde{k}\sim k_z\sim k_h N/\omega$. Putting these together, we place a very conservative upper bound of
\begin{equation}\label{eqn:Dw estimate}
    D_w\lesssim 7000\, d_c^2\,{\rm cm}^2/{\rm sec}.
\end{equation}
Note that $7000\,{\rm cm}^2/{\rm s}$ is similar to the microphysical chemical diffusivity expected in massive stars. However, for some wave interactions, we estimate $d_c\sim 100$, giving an upper bound corresponding to moderate enhancement of the chemical diffusivity.

However, theory and numerical experiments suggest the vertical velocities of waves excited by convection are much smaller than $\omega/k_z$. And of course, this theory of weakly nonlinear wave transport is not valid for such strongly nonlinear waves.
As an example, the wave luminosity spectrum measured in simulation of a $15M_\odot$ ZAMS model in \citet{Anders2023b}, when extrapolated to the most energetic and nonlinear waves ($\omega\sim 1/\tau_c$ and $\ell=1$), gives a vertical velocity $w\approx 5\times 10^{-3} \, \omega/k_z$. However, these most energetic waves are rapidly damped by radiative diffusion. The waves that can propagate a typical distance $\sim R_c$, the convection zone radius, have twice higher frequency, amplitude lower by a factor of $0.05$, and have $w\approx 10^{-4} \, \omega/k_z$. Hence, we expect the waves to be weakly nonlinear and for the transport to be well modeled by the calculations in this paper.
Since $D_w$ scales as the eighth power of the wave amplitude, even the most energetic waves with $w/(\omega/|k_z|)\approx5\times10^{-3}$ give a diffusivity smaller than eqn.~\ref{eqn:Dw estimate} by a factor of $(5\times10^{-3})^8$.

While this work has extensively tested this analytic framework for coherent 2D, Boussinesq, Cartesian gravity waves, it should also be tested in other systems that include more of the physics relevant in stellar interiors.
In terms of wave propagation, these include sphericity, effects of rotation and three-dimensionality, and extending beyond the Boussinesq equations (e.g., anelastic or fully compressible).
Although the detailed expressions derived here may differ when including these effects, we do not believe they will alter the main physical mechanisms beyond wave transport.
However, the treatment of superposition of multiple wave packets and the derivation of the final diffusion equation (\ref{eq:diffusion}) is more speculative, and should be tested with idealized modeling, as well as compared to more comprehensive numerical simulations \citep[e.g.,][]{Rogers2017,Herwig2023,Morton2025,Varghese2025}.
Upon verification, we believe it will be straightforward to implement this type of model of wave-driven chemical mixing into stellar evolution codes, to determine if it can help explain observed features of stars.

\begin{acknowledgments}

The authors would like to thank Greg Wagner for useful discussions regarding Stokes drift and Eulerian mean flows, as well as acknowledge useful discussions with Tami Rogers, Rathish Ratnasingham, Matteo Cantiello, Jim Fuller and William R. Young. We acknowledge the National Science Foundation and the Office of Naval Research for their support of the 2023 Woods Hole Oceanographic Institution Geophysical Fluid Dynamics Summer Program, where this work was initiated. YM is supported by a Scripps Institution of Oceanography Postdoctoral Fellowship. DL is partially supported by NSF AAG grant AST-2405812, Sloan Foundation grant FG-2024-21548 and Simons Foundation grant SFI-MPS-T-MPS-00007353.

\end{acknowledgments}





\newpage

\appendix

\section{Third-order tracer flux}  \label{append:proof}

In this Appendix, we show that the third-order tracer flux satisfies $\overline{ w_1 c_1^2 }=0$, which leads to \eqref{eq:c2s_old}. We decompose the first-order tracer perturbation into its slow and fast components $c_1=c_{1f}+c_{1s}$, where $c_{1s}$ is independent of the fast time $\tilde{t}$. From the first-order tracer equation, we have
\begin{equation}
    c_{1f} =
    -\frac{dc_0}{dz}\,\xi,
    \qquad
    \partial_{\tilde{t}}\xi=w_1,
\end{equation}
where $\xi$ is the vertical wave displacement. Expanding the cubic flux gives
\begin{equation} \label{append:flux}
    \overline{ w_1 c_1^2 } =
    \overline{ w_1 c_{1f}^2 } +
    2\overline{ w_1 c_{1f}c_{1s} } +
    \overline{ w_1 c_{1s}^2 }.
\end{equation}
Each term vanishes separately. Since $c_{1s}$ is independent of the fast time, the last term in \eqref{append:flux} is zero. The second term in \eqref{append:flux} follows 
\begin{align}
    \overline{w_1 c_{1f}c_{1s}} = -\frac{dc_0}{dz} \, c_{1s} \, \overline{\xi\partial_{\tilde{t}}\xi}
    = -\frac{1}{2}\frac{dc_0}{dz} \, c_{1s} \, \overline{\partial_{\tilde{t}}\xi^2}
    =0
\end{align}
by periodicity. 
Similarly, the third term in \eqref{append:flux} becomes
\begin{align}
    \overline{w_1 c_{1f}^2} = \left(\frac{dc_0}{dz}\right)^2
    \overline{\xi^2\partial_{\tilde{t}}\xi}
    = \frac{1}{3} \left(\frac{dc_0}{dz}\right)^2 \overline{\partial_{\tilde{t}}\xi^3}
    =0.
\end{align}

\bigskip
\section{Net vertical transport} \label{append:net}

We substitute the weakly nonlinear wave velocity field \eqref{eq:w1} into the vertical transport equation \eqref{eq:c2s}. Evaluating the resulting expressions symbolically using Mathematica, we find that the horizontally averaged slow-time tracer transport can be written in the form
\begin{equation}
    \begin{aligned} 
    \left\langle c_{2s}(z,t) \right\rangle = 
    \frac{\bar{t}^{2}}{2} \partial_z  & \Bigg[ \frac{d c_0}{\partial z} \sum_{i=1}^{n} \sum_{j=1}^{n}\sum_{k=1}^{n} \sum_{l=1}^{n}
    \delta_{\omega_i,\omega_j} \, \delta_{\omega_k,\omega_l}
    ~ e^{-z\left(\frac{1}{\mu_i}+\frac{1}{\mu_j}+\frac{1}{\mu_k}+\frac{1}{\mu_l}\right)}  A_i A_j A_k A_l  
    \\
    &\qquad   \Big\langle
    \widetilde{\alpha}_{1} \cos(\Delta_{+})
    + \widetilde{\alpha}_{2} \cos(\Delta_{-})
    + \widetilde{\alpha}_{3} \sin(\Delta_{+})
    + \widetilde{\alpha}_{4} \sin(\Delta_{-})
    \Big\rangle  \Bigg],
\end{aligned}
\end{equation}
where the phases are
\begin{align*} 
    \Delta_+ &= \left(k_{h,i}-k_{h,j}+k_{h,k}-k_{h,l}\right)x + \left(k_{z,i}-k_{z,j}+k_{z,k}-k_{z,l}\right)z + \varphi_i-\varphi_j+\varphi_k-\varphi_l, \\
    \Delta_- &= \left(k_{h,i}-k_{h,j}-k_{h,k}+k_{h,l}\right)x + \left(k_{z,i}-k_{z,j}-k_{z,k}+k_{z,l}\right)z + \varphi_i-\varphi_j-\varphi_k+\varphi_l.
\end{align*}
The transport coefficients $\widetilde{\alpha}_m=\widetilde{\alpha}_m(k_{h,\{i,j,k,l\}},k_{z,\{i,j,k,l\}},\mu_{\{i,j,k,l\}},\omega_{\{i,j,k,l\}},\kappa,N^2)$, $m=1,2,3,4$, are determined by the wave parameters and the background stratification.

To simplify the expression, we note that exchanging the indices $k$ and $l$ in the phase $\Delta_-$ maps $\Delta_-$ onto $\Delta_+$. The two phase contributions can therefore be combined into a single phase, yielding
\begin{equation}
    \begin{aligned} 
    \left\langle c_{2s}(z,t) \right\rangle &= 
    \frac{\bar{t}^{2}}{2} \partial_z \Bigg[\frac{d c_0}{\partial z} \sum_{i=1}^{n} \sum_{j=1}^{n}\sum_{k=1}^{n}\sum_{l=1}^{n}
    \delta_{\omega_i,\omega_j} \, \delta_{\omega_k,\omega_l}
    ~ e^{-z\left(\frac{1}{\mu_i}+\frac{1}{\mu_j}+\frac{1}{\mu_k}+\frac{1}{\mu_l}\right)}   A_i A_j A_k A_l
    \langle
    \alpha_1 \cos(\Delta)
    + \alpha_2 \sin(\Delta)
    \rangle \Bigg],
\end{aligned}
\end{equation}
where
\begin{align*}
    \Delta &= \left(k_{h,i}-k_{h,j}+k_{h,k}-k_{h,l}\right)x + \left(k_{z,i}-k_{z,j}+k_{z,k}-k_{z,l}\right)z + \varphi_i-\varphi_j+\varphi_k-\varphi_l, \\
    \alpha_1 &= \widetilde{\alpha}_1+\widetilde{\alpha}_2 \big|_{k \leftrightarrow l}, \qquad
    \alpha_2 = \widetilde{\alpha}_3+\widetilde{\alpha}_4 \big|_{k \leftrightarrow l}.
\end{align*}
We therefore obtain the expression \eqref{eq:c2s_n}.

\bibliography{main}{}
\bibliographystyle{aasjournalv7}

\end{document}